\documentclass[12pt,preprint]{aastex}

\shorttitle{The Astrophysical Behavior of Open Clusters along the Milky Way Galaxy}
\shortauthors{Tadross, A. L.}

\begin{document}

\title{The Astrophysical Behavior of Open Clusters along the Milky Way Galaxy}

\author{Tadross, A. L.}
\affil{National Research Institute of Astronomy and Geophysics, 11421 - Helwan, Cairo, Egypt.}
\email{altadross@yahoo.com}

\begin{abstract}
The main aim of this paper is to study the astrophysical behaviour of open clusters' properties along the Milky Way Galaxy. Near-IR {\it JHK$_{S}$} {\it (2MASS)} photometry has been used for getting a homogeneous Catalog of 263 open clusters' parameters, which are randomly selected and studied by the author through the last five years; most of them were studied for the first time. The correlations between the astrophysical parameters of these clusters have been achieved by morphological way and compared with the most recent works.
\end{abstract}


\section{Introduction}

Open star clusters are very important objects in solving problems of star formation, stellar evolution, and improving our knowledge about the distance scale and the kinematic properties of the Milky Way Galaxy. This kind of study requires a large set of homogeneous data on the positions and ages of open star clusters, which are estimated in a precision way from their Colour-Magnitude Diagrams {\it (CMDs)}.
However, it is useful to re-investigate the properties and structures of the Milky Way Galaxy using the most recent Near--IR {\it JHK$_{S}$} photometric data from the 2--Micron All Sky Survey $(\it {2MASS})$ Point Source Catalogue of Skrutskie et al. (2006). In this context, we used a sample of 263 open star clusters (most of them are studied for the first time) that have been analysed by the author, in a series of papers; Tadross (2008 $\sim$ 2012); see Tables 1 and 2. Our aim is to repeat the work we have done 12 years ago, Tadross 2001 and Tadross et al. 2002, to study such relations using NIR observations instead of UBV.

This paper is organized as follows. In Sec. 2, a historical review of the present study is obtained. In Sec. 3, data analysis of the clusters under investigation are presented. The limiting and core radii are given in Sec. 4. The main photometric parameters are obtained in Sec. 5. Ages and locations, distribution are presented in Sec. 6. Reddening distribution is presented in Sec. 7. The diameters and ages' relations are given in Secs. 8 and 9 respectively. Secs. 10 and 11 describe the spiral arms and warp of the Galaxy respectively. The conclusions are obtained in Sec. 12.

\section{Historical Review}

Recently, Bukowiecki et al. (2011) determined new coordinates of the centres, angular sizes and radial density profiles for 849 open clusters in the Galaxy based on the {\it 2MASS} database. Froebrich et al. (2010) studied 269 open clusters; ages, core radii, reddening, Galactocentric distances and the scale-heights were determined. Similar to this kind of study, Schilbach et al. (2006) derived the linear sizes of some 600 clusters and investigated the effect of the mass segregation of stars in open clusters. Bica et al. (2003) researched 346 open clusters, based on the 2MASS database; they studied the linear diameters and spatial distribution of open clusters in the Galaxy. Tadross (2001) and Tadross et al. (2002) studied 160 open clusters used {\it UBV-CCD} observations and derived the relationships projected onto the Galactic plane in morphological way. Dutra \& Bica (2001) studied 42 new infrared star clusters, stellar groups and candidates towards the Cyngus X region. Dutra \& Bica (2000) have studied 103 Galactic open clusters and compared the reddening values obtained from far infrared {\it IRAS} and {\it COBE} observations with those obtained from visible observations. Dambis (1999) determined the main parameters of 203 open clusters based on the published photoelectric and CCD data. Malysheva (1997) published a Catalogue of parameters for 73 open clusters determined from uvby$\beta$ photometry; his values are in good agreement with those of Loktin \& Matkin (1994). Loktin et al. (1997) published his improved version Catalogue, which contained the updated parameters of homogeneously estimated excesses, distances, and ages for 367 open clusters. Friel (1995) and Janes \& Phelps (1994) based on a sample of some 70 objects investigated how the extinction and age depend on the position in the Galaxy. Also, a comparison between the age and position in the Galaxy was studied by Lyng{\aa} (1980, 1982). In addition an old study of Janes (1979) used UBV photometry to study the reddening and metallicity of 41 open clusters.

\section{Data analysis}

The current study depended mainly on the correlations between the astrophysical parameters of 263 open star clusters of different names listed as follows: 124 clusters of NGC;  24 objects of Berkeley; 23 of Kronberger; 23 of Czernik; 11 of Dol-Dzim; 11 of Ruprecht; 10 of Dolidze; 6 of Dias; 5 of Turner; 4 of King; 3 of BH; 3 of Eso; 3 of IC; 2 of Alessi; 2 of Juchert; 2 of Riddle; 2 of Skiff; 2 of Teutsch; 1 of Collinder; 1 of Patchick; and 1 cluster of Toepler.
\\
This sample contains clusters with ages in the range from 5 Myr to 5 Gyr. They are located at distances up to 4.7 kpc from the Sun ($R_{_{\odot}}$), up to 12.5 kpc from the Galactic Centre ($R_{gc}$), and less than $\pm$ 2 kpc from the Galactic Plane ($Z$). They range from 0.25 to 16.5 arcmin in limiting radii ($R_ {lim.}$), and up to 1  arcmin in core radii ($R_{c}$); with noticing that the estimated $R_ {lim.}$ and $R_{c}$ in parsecs depending mainly on the distance of the clusters individually.
\\ \\
Data extraction has been performed for each cluster using the known tool of VizieR for {\it 2MASS} \footnote{\it http://vizier.u-strasbg.fr/viz-bin/VizieR?-source=2MASS} Point Source Catalogue database of Skrutskie et al. (2006). The investigated clusters have been selected from WEBDA and DIAS databases under some conditions mentioned in our last series papers. It is noticed that most clusters' sizes seem to be greater in the infrared band (2MASS observations) than in the optical band; because this system can detect the very faint stars, even those behind the curtains of interstellar matter. The real spatial distribution of open clusters along the Milky Way Galaxy refers to some paucity of the clusters at G. longitudes range from $140^{o}$ to $200^{o}$. The lack of objects in that direction is noticed also in earlier studies and confirmed for open clusters by Benjamin (2008), and Bukowiecki et al. (2011). Older clusters seem to be more dispersed than younger ones of the Hyades, as shown in the left panel of Fig. 1. The relationships between the astrophysical parameters of open clusters are presented here with respect to their ages and places, so then the astrophysical behaviour of open clusters along the Milky Way Galaxy can be investigated.


\section{Limiting and core radii}
One of the main tasks in this work was the determination of the radial density profile (RDP) for each cluster, i.e. the observed stellar density $\rho$ that plotted as a function of the angular radial distance from the cluster centre, King (1966):
\begin{center}
$\rho(r)=f_{bg}+\frac{f_{0}}{1+(R_{lim}/R_{c})^{2}}$
\end{center}
where $R_{c}$, $f_{0}$, and $f_{bg}$ are the core radius, the central density, and the background density, respectively. The core radius was derived as a distance where the stellar density drops to half of $f_{0}$. The parameters were derived with the least--square method. The cluster's limiting radius, $R_ {lim} $, was defined by comparing $\rho(r)$ with the background density level $\rho_{bg}$, (cf. Bukowiecki et al. 2011).
From both $R_{c}$ and $R_{lim}$, one can estimate the concentration parameter $c=log ({R_{lim}}/{R_{c}})$, Peterson \& King (1975). This parameter can be added as a new item to characterise the structure of clusters along the Galaxy. In the present work, the concentration parameters are ranging from 0.39 to 2.5. In this context, Nilakshi et al. (2002) concluded that the angular size of the coronal region is about 6 times the core radius, while Maciejewski \& Niedzielski (2007) reported that $R_{lim}$ may vary for individual clusters from 2$R_{c}$ to 7$R_{c}$. In our case, for the whole sample, the average values of limiting radius, core radius, and concentration parameter are 4.6 arcmin, 0.3 arcmin, and 1.2 respectively. We concluded that $R_{lim}$= 6.85 $R_{c}$; for the clusters up to $R_{c}$= 0.5 arcmin, and $R_{lim}$= 2.88 $R_{c}$; for the clusters up to $R_{c}$ = 1.0 arcmin. i.e, our conclusion is almost in agreement with Maciejewski \& Niedzielski (2007).

\section{Main photometric parameters}
Depending on the 2MASS data, deep stellar analyses of the candidate clusters have been presented. The photometric data of 2MASS not only allow us to construct of relatively well defined CM diagrams of the clusters, but also permit a more reliable determination of astrophysical parameters. In this paper, we used extraction areas having a radius of 20 arcmin, which are larger than the estimated limiting radius of the clusters. Because of the weak contrast between the cluster and the background field density, some inaccurate statistical results may be produced beyond the real limit of cluster borders (Tadross, 2005).

The main astrophysical parameters of the clusters, e.g. age, reddening, distance modulus, can be determined by fitting the isochrones to the cluster CMDs. To do this, we applied several fittings on the CMDs of the clusters by using the stellar evolution models of Marigo et al. (2008) of Padova isochrones on the solar metallicity. It is worth mentioning that the assumptions of solar metallicity are quite adequate for young and intermediate age open clusters, which are closer to the Galactic disk. So, Near-Infrared surveys are very useful for the investigation of such clusters. It is relatively less affected by high reddening from the Galactic plane. However, for a specific age isochrones, the fit should be obtained at the same distance modulus for both diagrams [J-(J-H) \& $K_{s}$-(J-$K_{s}$)], and the color excesses should obey Fiorucci \& Munari (2003)'s relations for normal interstellar medium. We note that, it is difficult to obtain some accurate determinations of the astrophysical parameters due to the weak contrast between clusters and field stars.
\\
Reddening determination is one of the major steps in the cluster compilation. Therefore, it estimated guiding by Schlegel et al. (1998) in our estimations. In this context, for color excesses transformations, we used the coefficient ratios $\frac {A_{J}}{A_{V}}=0.276$ and $\frac {A_{H}}{A_{V}}=0.176$, which were derived from absorption rations in Schlegel et al. (1998), while the ratio $\frac {A_{K_s}}{A_{V}}=0.118$ was derived from Dutra et al. (2002). Applying the calculations of Fiorucci \& Munari (2003) for the color excess of {\it 2MASS} photometric system; we ended up with the following results: $\frac {E_{J-H}}{E_{B-V}}=0.309\pm0.130$, $\frac {E_{J-K_s}}{E_{B-V}}=0.485\pm0.150$, where R$_{V}=\frac {A_{V}}{E_{B-V}}= 3.1$. Also, we can de-reddened the distance modulus using these formulae:  $\frac {A_{J}}{E_{B-V}}$= 0.887, $\frac {A_{K_s}}{E_{B-V}}$= 0.322. Then the distance of each cluster from the Sun, $R_{\odot}$, can be calculated. Consequently, the distance from the Galactic plane ($Z_{\odot}$), and the projected distances in the Galactic plane from the Sun ($X_{\odot}~\&~Y_{\odot}$) can be determined, see Table 3. For more details about the distance calculations, see Tadross (2011).

\section{Ages and locations}
The distribution of our sample according the distances from the Galactic center, Rgc, and the height from the Galactic plane, $Z$, is presented in the right panel of Fig. 1. We can see that the clusters with ages younger than Hyades, i.e. less than $(7\times10^{{\rm 8}} yr)$ are strongly concentrated to the Galactic plane. While the clusters which are older than Hyades are more dispersed from the Galactic plane (cf. Friel 1995). However, the correlations of the clusters' ages and locations with the other properties along the Milky Way Galaxy are presented in the following sections.

\section{Reddening distribution}

In fact reddening affects the distance determination via the main sequence fitting, actually it affected all the cluster's dimensions and positions on the Galaxy (cf. Tadross et al. 2002). The distribution of the reddening of our sample versus the Galactic latitudes confirms that the higher values of reddening are concentrated on and near the Galactic plane as shown in Fig. 2. Along the Galactic longitude bins of 20$^{\circ}$ the distribution of the mean reddening at each bin show that the higher values are concentrated around the longitude range from 345$^{\circ}$ to 130$^{\circ}$, i.e. in the directions of the Galactic centre and Sagittarius arm, as shown in Fig. 3. The general trend of reddening with age shows that reddening decreases with ages where younger clusters tend to be more reddened than older ones, see Fig. 4. On the other hand, the relation between reddening and $R_{gc}$ reveals that the clusters inside the galactocentric radius of the Sun ($R_{gc_{\odot}} = 8.5$ kpc) have higher values of reddening than that of outside ones, as shown in Fig. 5. This confirms to some extent that the Sun's vicinity clusters are young and medium ones than those  outside clusters.

\section{Diameters' relations}

The linear diameters have been plotted versus the absolute values of the height from the Galactic plane $\vert $Z$\vert $, and the distance from the Galactic centre $R_{gc}$ as shown in Fig. 6 and Fig. 7 respectively. We can see that most clusters with typical diameters ($D<10$ pc) are concentrated near the Galactic plane, especially those inside the galactocentric radius of the Sun $R_{gc}\leq 8.5$ kpc. The general trend of this relation appears that large, old clusters are found far from $R_{gc}$ - it is confirmed by Bukowiecki et al. (2011)- and also at large height $Z$ (Tadross et al. 2002). There are some young clusters with larger diameters belong to the Galactic plane are loose and unbound objects.
The relation between the diameters and the galactocentric radii has been examined to be:
\begin{center}
Diam. = 0.53 $R_{gc}$ - 0.19
\end{center}
The standard error of this relation $\approx$ 3.0\\
Burki and Maeder (1976) found a correlation between these quantities only for the very young clusters, but we have found such correlation for intermediate and older clusters as well.
\\ \\
Fig. 8 represents the relation between ages and linear diameters of our sample. It can be expressed as follows:
\begin{center}
Diam. =  3.18 \emph{Log (age)} - 18.53
\end{center}
The standard error of this relation $\approx$ 3.2\\
We can see that, to some extent, there is a correlation between diameters and ages, whereas clusters of large sizes belong to older ages, which have also large heights from $Z$. Youngest clusters with large sizes are supposed to be some groups of OB associations and probably they are not bound systems (Lyng{\aa} 1982; Janes, Tilley \& Lyng{\aa} 1988). Massive clusters with small sizes will be dissolved due to encounters among their members, while those of very large sizes with the same mass will be unstable in the Galactic tidal field, and they may take very long time to have stability and relaxation (Theis 2001). Small clusters with typical diameters less than 10 pc show a concentration to the Galactic plane (Wielen 1971 and 1975) in the range of $\vert Z\vert <100$ pc, see Fig. 6, but larger clusters have both intermediate and old ages.

\section{Ages' relations}

The cluster's ages has been plotted versus $Z$, as shown in Fig. 9. Most clusters with ages $t\leq10^{{\rm 8.9}} $ yr is lying around $\vert Z\vert\approx$ 200 PC, while older ones are lying higher than such heights. It indicated that the thickness of the Galactic disk has not changed on the time scale of about $10^{{\rm 9.0}}$ yr and the clusters can be formed everywhere inside this layer (cf. Tadross et al. 2002). Lyng{\aa} \& Palous (1987) have found that old clusters are much thicker distributed in the outer parts of the Galaxy than the inner parts, Bukowiecki et al. (2011). Also in our study the thickness of the Galactic disk  increases for older clusters as well. Old clusters not only spend their time in the outer disk away from the disruptive effects of giant molecular clouds, but also, they spend their time at large distances from the Galactic plane, further enhancing their survivability (Friel 1995).
\\
On the other hand, the relation between ages and $R_{gc}$ of the clusters implies that there is a lack of old clusters in the inner parts of the Galactic disk, and the anti-center clusters survive longer than such clusters. In the inner parts of the Galaxy they have never gotten the relaxation state in the fluctuating gravitational field of that part (Lyng{\aa}, 1980; McClure et al., 1981; Vanden Bergh, 1985). The general trend reveals that, lifetime increases outwards the Milky Way Galaxy, where clusters live longer than those in the inner parts of it.

\section{Galactic spiral arms}
To show the shape of the spiral arms of the Galaxy, several studies have been carried out in the last five decades. The positions of the clusters on the Galactic plane have been used to trace the spiral arms of the Milky Way Galaxy. Centered on the Sun at ($X_{\odot}$ = $Y_{\odot} = 0$) the distribution of the clusters on the Galactic plane has been plotted, as shown in Fig. 10. Within a radius of 4 kpc from the sun (cf. Jeanes, Tilley \& Lyng{\aa} 1988) the distribution of the studied clusters define three concentration features which are related to the spiral structure of the Galaxy, i.e. Perseus, Sagittarius and Carina. It is assumed that there are more than three arms of the Galaxy but because of the patchy cloud and absorption effects we can't able to detect them all!.

\section{Galactic warp}
The effect of the Galactic warp may be declared from the distribution of the open clusters of our sample using the Galactic coordinates $X$ and $Y$ versus the height $Z$ within $\pm $2 kpc from the Galactic plane, as shown in Fig. 11. The directions of the G. X and G. Y defined to be positive in the direction of the Galactic radial center and towards the direction of Galactic rotation respectively (Jeanes, Tilley \& Lyng{\aa} 1988; Camer\'{o}n 1999; Piatti et al. 2003). No strong indication to the warp has been detected on the G. X direction, but, to some extent, it can be detected on the G. Y direction. This may refer to the leak of the studied clusters, especially those have large distances from the sun's vicinity (Tadross et al. 2002).

\section{Conclusion}

The results of our studied clusters in the last five years using near--infrared {\it JHK$_{S}$} photometric system are obtained here, and the correlations between the astrophysical parameters along the Milky Way Galaxy are achieved. It is obvious that ({\it JHK$_{S}$}) 2MASS system affected the magnitude limit of the clusters, which detects many faint members located away from the cluster's core, so then the cluster seems to be larger than in optical bands. Detecting stars located in the lower parts of CMDs, make the fitting with standard zero age main sequence much easier. This, of course, has contributed to the evaluation of the cluster parameters, i.e. distances, diameters, ages, reddening, etc. From our reduction, we concluded that $R_{lim}$= 6.85 $R_{c}$; for the clusters up to $R_{c}$= 0.5 arcmin, and $R_{lim}$= 2.88 $R_{c}$; for the clusters up to $R_{c}$= 1.0 arcmin, which are in agreement with Maciejewski \& Niedzielski (2007). We also noticed that the linear size of open clusters increases with ages. The reddening decreases outward the Galactic plane, $Z$ and the Galactic Center, $R_ {gc} $, as well. This is noticed also for clusters located near the Sun vicinity and further than 8.5 kpc from the Galactic centre, i.e. the density of dust and gas decreases, too.
\\ \\
From our analysis, we noticed that the number of clusters decreases with $Z$; more than half of the studied clusters ($52\%$) have aged less than 500 mega years and located at average $|Z|$ = 75 pc. Hence, the older ones are located at average $|Z|$ = 275 pc, which is in agreement with Bukowiecki et al. (2011). We can show that the difference between younger and older clusters can be declared in locations and sizes as the following relation:
\begin{center}
Diam. = 0.53 $R_{gc}$ - 0.19 =  3.18 \emph{Log (age)} - 18.53
\end{center}
We found that the number of older clusters increases with $R_{gc}$ and younger ones are obtained at an average $R_{gc} = 8.8$ kpc, which is confirmed by Tadross et al. (2002), Froebrich (2010), and Bukowiecki et al. (2011). The paucity of the clusters at G. longitudes range from $140^{o}$ to $200^{o}$ is noticeable by Tadross et al. (2002), Benjamin (2008), Froebrich (2010), and Bukowiecki et al. (2011). It may reflect the real spatial structure of the Milky Way Galaxy in that direction near the feature region of the Perseus arm (the external youngest arm of the Galaxy).

\begin{acknowledgements}
This publication makes use of data products from the Naval Observatory Merged Astrometric Dataset {\it (NOMAD)} and the Two Micron All Sky Survey {\it 2MASS}, which is a joint project of the University of Massachusetts and the Infrared Processing and Analysis Centre/California Institute of Technology, funded by the National Aeronautics and Space Administration and the National Science Foundation. Catalogues from {\it CDS}/{\it SIMBAD} (Strasbourg), and Digitized Sky Survey {\it DSS} images from the Space Telescope Science Institute have been employed.
\end{acknowledgements}

\begin{figure*}
\begin{center}
      {\includegraphics[width=16cm]{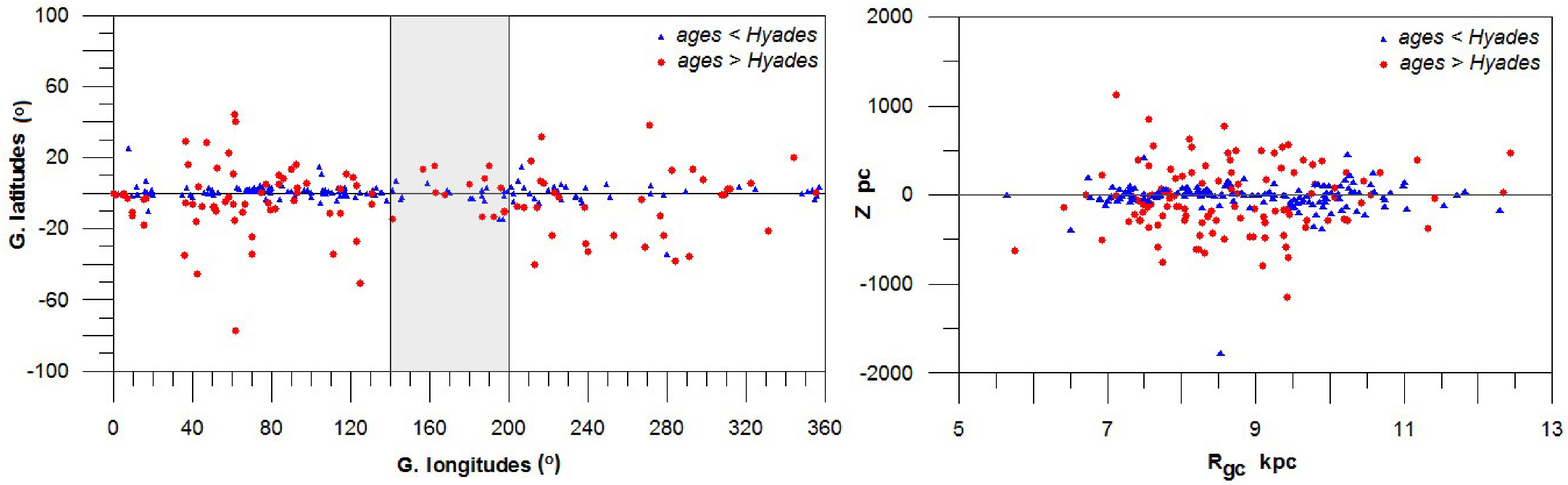}}
      \end{center}
      \caption{Left panel represents the clusters' distribution according to their Galactic longitudes and latitudes, the dark area refers to the paucity of the clusters at G. longitudes range from $140^{o}$ to $200^{o}$. Right panel represents the clusters' distribution according their distances from the Galactic centre, $R_{gc}$, and Galactic plane, $Z$, assuming that $Z_{_{\odot}}$ = -33 pc, and $R_{gc_{\odot}}$ = 8.5 kpc for the Sun. Both panels plotted for two ranges of clusters' ages, younger and older than Hyades.}
\end{figure*}

\begin{figure}
\begin{center}
      {\includegraphics[width=12cm]{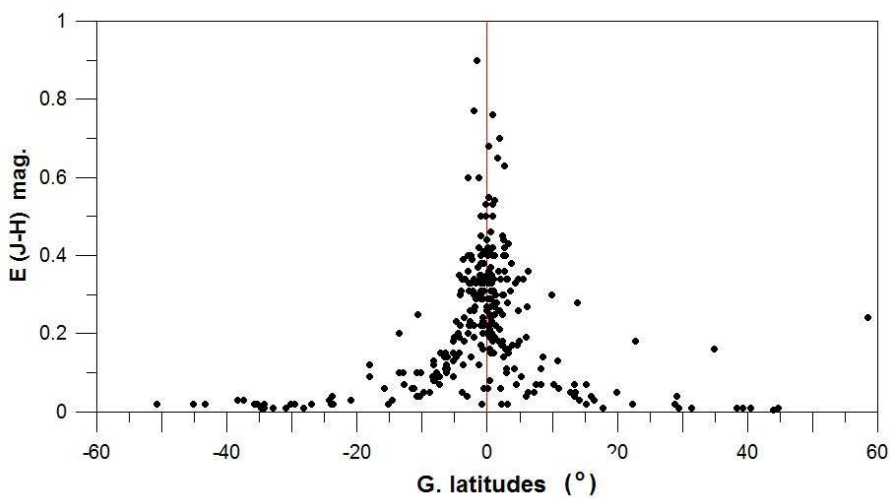}}
      \end{center}
      \caption{The distribution of the reddening versus the Galactic latitudes.}
\end{figure}

\begin{figure}
\begin{center}
      {\includegraphics[width=11cm]{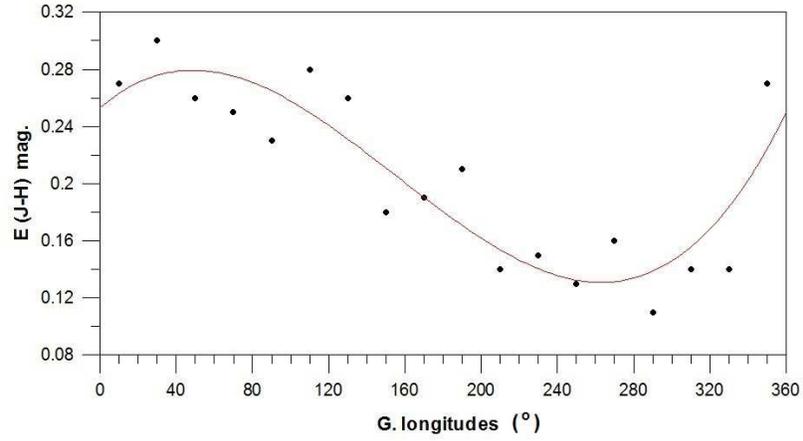}}
      \end{center}
      \caption{The distribution of the mean reddening along the Galactic longitude bins of 20 degrees for each.}
\end{figure}

\begin{figure}
\begin{center}
      {\includegraphics[width=11cm]{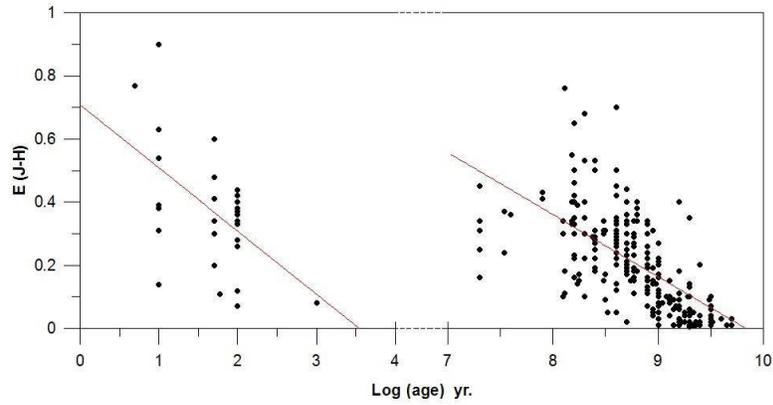}}
      \end{center}
      \caption{The relation between the clusters' ages and reddening. Left and right panels represent the very young and old clusters respectively.}
\end{figure}

\begin{figure}
\begin{center}
      {\includegraphics[width=11cm]{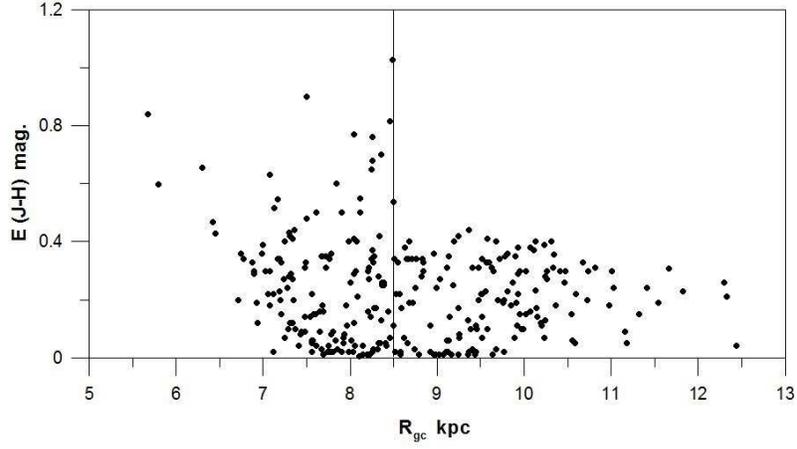}}
      \end{center}
      \caption{The relation between the reddening and distance from the Galactic centre $R_{gc}$. The vertical line represents the galactocentric radius of the Sun, assuming that $R_{gc_{\odot}}$ = 8.5 kpc.}
\end{figure}

\begin{figure}
\begin{center}
      {\includegraphics[width=11cm]{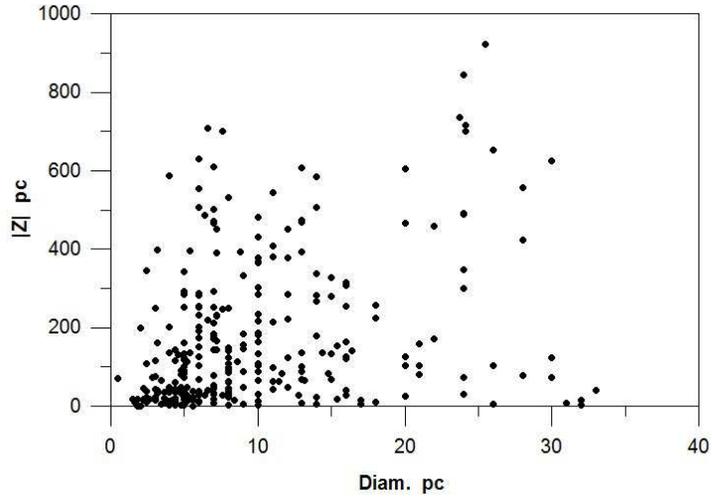}}
      \end{center}
      \caption{The relation between the diameters and the absolute values of the height from the Galactic plane, $\vert $Z$\vert $.}
\end{figure}

\begin{figure}
\begin{center}
      {\includegraphics[width=10cm]{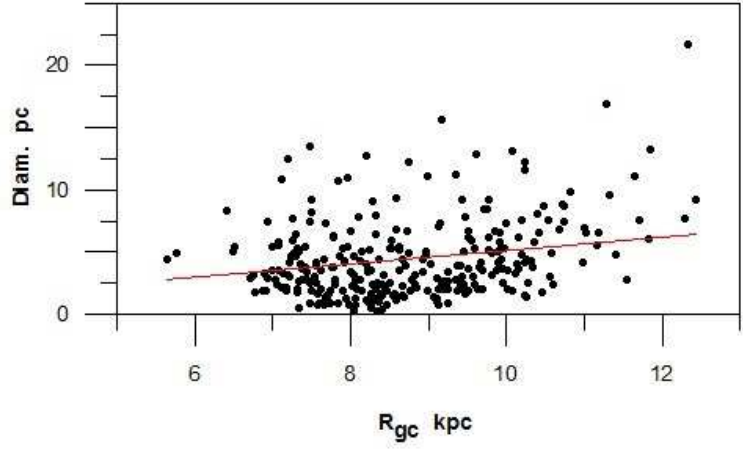}}
      \end{center}
      \caption{The relation between the galactocentric radii, $R_{gc}$, and linear diameters of the clusters, assuming that $R_{gc_{\odot}}$ = 8.5 kpc for the Sun. The standard error $\approx$ 3.0}.
\end{figure}

\begin{figure}
\begin{center}
      {\includegraphics[width=10cm]{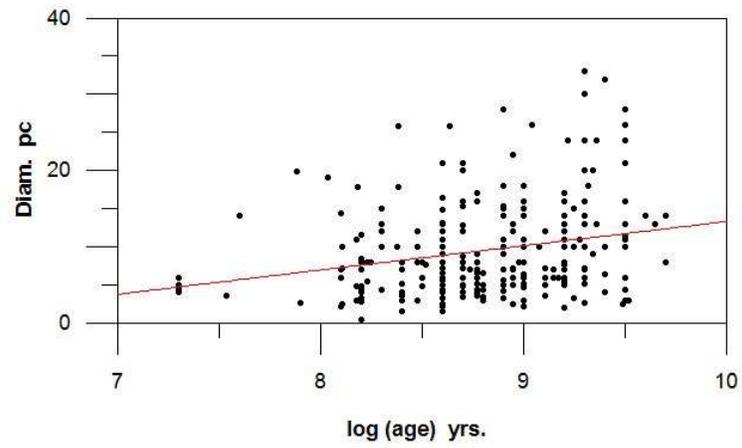}}
      \end{center}
      \caption{The relation between ages and linear diameters of the studied clusters. The standard error $\approx$ 3.2.}
\end{figure}

\begin{figure}
\begin{center}
      {\includegraphics[width=11cm]{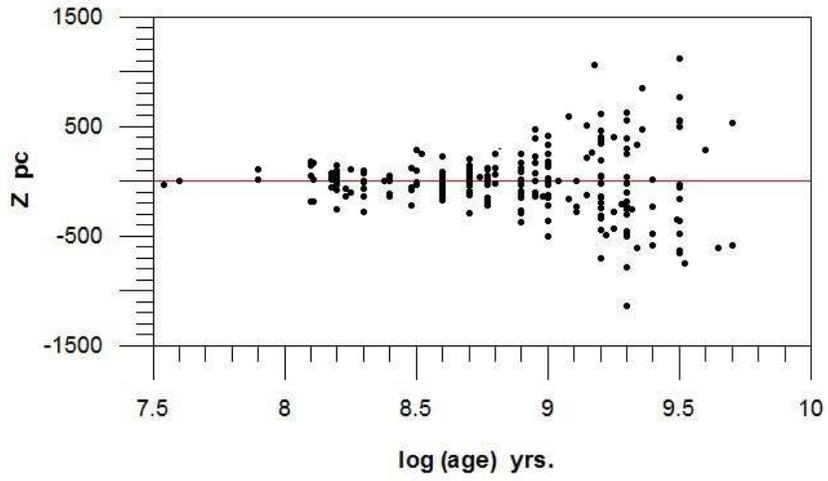}}
      \end{center}
      \caption{The relation between ages and the heights from the Galactic plane, assuming that $Z_{_{\odot}}$ = -33 pc for the Sun.}
\end{figure}

\begin{figure}
\begin{center}
      {\includegraphics[width=10cm]{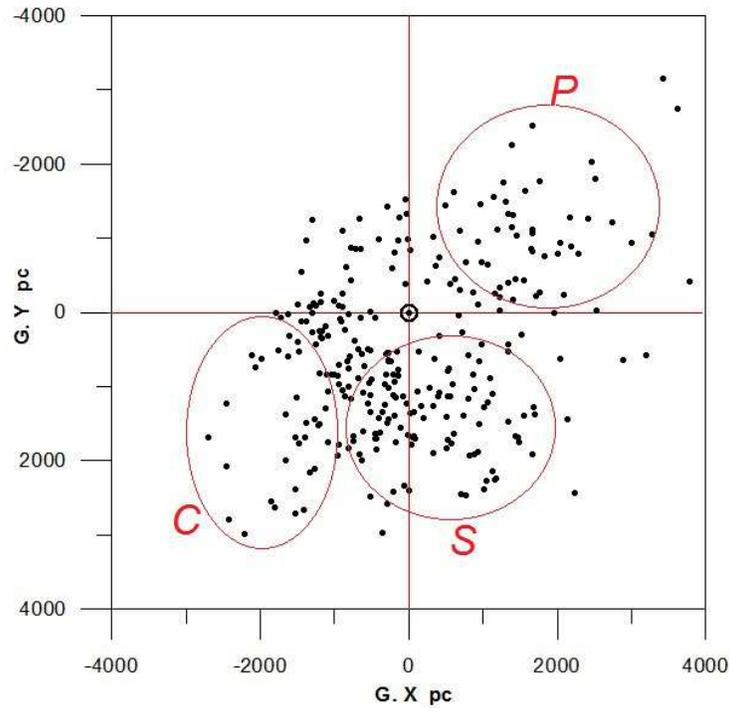}}
      \end{center}
      \caption{The distribution of our sample on the Galactic plane. The three circles refer to the three famous arms of the Galaxy: Perseus (P), Sagittarius (S), and Carina (C).}
\end{figure}

\newpage
\begin{figure*}
\begin{center}
      {\includegraphics[width=14cm]{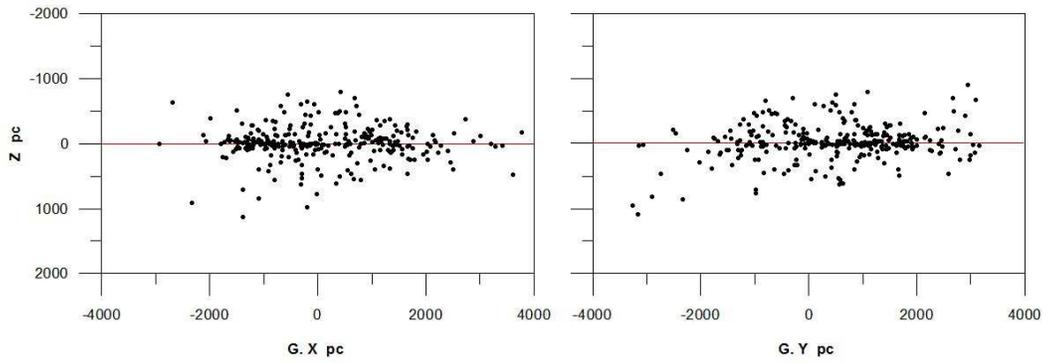}}
      \end{center}
      \caption{The distribution of our sample according to their Galactic coordinates $X_{\odot}$ and $Y_{\odot}$ versus the height from the Galactic plane $Z$. The Sun's position is at $X_{\odot}$= $Y_{\odot}$ = 0.}
\end{figure*}

\begin{table*}
\caption{The astrophysical main parameters for the studied clusters, derived by the author. Columns display, respectively, cluster name, equatorial positions, angular limiting radius, core radius, age, reddening, distance from the sun $R_{\odot}$, distance from the Galactic centre $R_\mathrm{gc}$, the projected distances on the Galactic plane from the sun $X_{\odot}$, $Y_{\odot}$, $Z$, and the reference for each cluster.}

\centerline{
\begin{tabular}{cccrrrrrrrrrr}
\hline
            Cluster  & $\alpha$
            & $\delta$ & R$_{lim.}$ & R$_{c}$& Age  & E$_{B-V}$  & $R_{\odot}$  & $R_\mathrm{gc}$
            & $X_{\odot}$  & $Y_{\odot}$ & $Z$ & Ref.\\
            & $~~^{h}~~^{m}~~^{s}$ & $~~^{\circ}~~{'}~~{''}$ & ${'}$ & ${'}$ & \it Gyr & \it mag  & \it pc  & \it kpc   & \it pc
            & \it pc  & \it pc & \\
\hline
NGC ~110 &00 27	25 &71	23  00 &11.0 &0.60 & 0.9 $^{\pm0.04}$ & 0.46 $^{\pm0.10}$ & 1150 $^{\pm53}$ & 9.14  & 585   & 975   & 172 & 6~~~\\
NGC ~272 &00 51	24 &35	49	54 &~3.2 &0.29 &2.5 $^{\pm0.10}$ & 0.06 $^{\pm0.02}$ & 1068 $^{\pm50}$ & 9.12  & 517   & 798   & --486 & 6~~~\\
NGC ~657 &01 43	21 &55	50	11 &~3.0 &0.90 &1.6 $^{\pm0.11}$ & 0.34 $^{\pm0.05}$ & 1372 $^{\pm63}$ & 9.44  & 881   & 1041  & --151 & 6~~~\\
NGC ~743 &01 58	37 &60	09	18 &~4.0 &0.26 &0.5 $^{\pm0.02}$ & 0.95 $^{\pm0.20}$ & 1618 $^{\pm75}$ & 9.64  & 1065  & 1217  & --46 & 6~~~\\
NGC ~956 &02 32	30 &44	35	36 &5.0 &0.64 &1.0 $^{\pm0.04}$ & 0.10 $^{\pm0.05}$ & 1455  $^{\pm67}$ & 9.68  & 1097  & 883    & --367 & 6~~~\\
NGC 1498 &04 00	18 &-12	00	54 &~3.8 &0.68 &1.6 $^{\pm0.11}$ & 0.04 $^{\pm0.02}$ & 1020 $^{\pm47}$ & 9.44  & 680   & --297  & --700 & 6~~~\\
NGC 1520 &03 57	51 &-76	47	42 &~3.6 &0.50 &2.0 $^{\pm0.08}$ & 0.06 $^{\pm0.01}$ & ~775 $^{\pm36}$ & 8.25  & --227 & --587  & --452 & 6~~~\\
NGC 1557 &04 13	11 &-70	28	18 &13.0 &0.10 &3.0 $^{\pm0.12}$ & 0.11 $^{\pm0.05}$ & 1055  $^{\pm49}$ & 8.31  & --197 & --805  & --653 & 6~~~\\
NGC 1724 &05 03 32 &49	29	30 &~8.0 &0.80 &0.6 $^{\pm0.02}$ & 0.57 $^{\pm0.10}$ & 1437  $^{\pm66}$ & 9.85  & 1332  & 526    & 121 & 6~~~\\
NGC 1785 &04 58	35 &-68	50	40 &~3.0 &0.25 &0.5 $^{\pm0.02}$ & 0.06 $^{\pm0.02}$ & 3080 $^{\pm140}$& 8.52  & --437 & --2477 & --1777 & 6~~~\\
NGC 1807 &05 10	43 &16	31	18 &~9.0 &0.46 & 1.0   $^{\pm0.04}$ & 0.32 $^{\pm0.05}$ & ~960  $^{\pm44}$ & 9.46  & 928   & --99   & --224 & 6~~~\\
NGC 1857 &05 20	12 &39	21	00 &~4.0 &0.08 &0.16 $^{\pm0.12}$ & 0.97 $^{\pm0.20}$ & 1545  $^{\pm71}$ & 10.02 & 1513  & 310   & 34  & 6~~~\\
NGC 1891 &05 21	25 &-35	44	24 &10.0 &0.14 &2.0 $^{\pm0.05}$ & 0.03 $^{\pm0.02}$ & ~860  $^{\pm40}$ & 8.96  & 364   & --624  & --467 & 6~~~\\
NGC 2013 &05 44	01 &55	47	36 &~3.0 &0.32 &1.5 $^{\pm0.06}$ & 0.23 $^{\pm0.05}$ & 1100 $^{\pm51}$ & 9.52  & 981   & 426    & 255 & 6~~~\\
NGC 2017 &05 39	17 &-17	50	48 &~6.0 &0.69 &1.6 $^{\pm0.11}$ & 0.06 $^{\pm0.02}$ & 1120  $^{\pm52}$ & 9.37  & 767   & --681  & --450 & 6~~~\\
NGC 2026 &05 43	12 &20	08	00 &~8.0 &0.38 &0.55 $^{\pm0.02}$ & 0.60 $^{\pm0.10}$ & 1418 $^{\pm65}$ & 9.91  & 1401  & --178  & --125 & 6~~~\\
NGC 2039 &05 44	00 &08	41	30 &~5.0 &0.36 &1.2 $^{\pm0.05}$ & 0.32 $^{\pm0.05}$ & ~920  $^{\pm42}$ & 9.38  & 863   & --273  & --164 & 6~~~\\
NGC 2061 &05 42	42 &-34	00	34 &~9.0 &0.96 &2.1 $^{\pm0.08}$ & 0.03 $^{\pm0.01}$ & ~542  $^{\pm25}$ & 8.79  & 246   & --409  & --256 & 6~~~\\
NGC 2063 &05 46	43 &08	46	54 &~6.0 &0.80 &1.3 $^{\pm0.05}$ & 0.32 $^{\pm0.05}$ & 1525  $^{\pm70}$ & 9.97  & 1430  & --445  & --285 & 6~~~\\
NGC 2132 &05 55	18 &-59	54	36 &12.0 &0.38 &1.65 $^{\pm0.12}$ & 0.06 $^{\pm0.02}$ & ~974 $^{\pm45}$ & 8.58  & 19    & --842  & --490 & 6~~~\\
NGC 2165 &06 11	04 &51	40	36 &~5.0 &0.44 &1.5 $^{\pm0.06}$ & 0.23 $^{\pm0.05}$ & 1445  $^{\pm67}$ & 9.89  & 1328  & 427    & 377 & 6~~~\\
NGC 2189 &06 12	09 &01	03	54 &~7.0 &0.78 &0.8 $^{\pm0.03}$ & 0.39 $^{\pm0.05}$ & 1869  $^{\pm86}$ & 10.19 & 1641  & --853  & --268 & 6~~~\\
NGC 2219 &06 23	44 &-04	40	36 &~3.5 &0.30 &0.8 $^{\pm0.03}$ & 0.40 $^{\pm0.08}$ & 2023 $^{\pm93}$ & 10.24 & 1660  & --1118 & --292 & 6~~~\\
NGC 2220 &06 21	11 &-44	45	30 &~6.5 &0.40 &3.0 $^{\pm0.12}$ & 0.06 $^{\pm0.01}$ & 1170  $^{\pm54}$ & 8.92  & 322   & --1020 & --475 & 6~~~\\
NGC 2224 &06 27	28 &12	35	36 &~7.0 &0.44 &0.01 $^{\pm0.00}$ & 1.00 $^{\pm0.25}$ & 2415 $^{\pm111}$& 10.81 & 2284  & --785  & 23 & 6~~~\\
NGC 2234 &06 29	24 &16	41	00 &14.0 &0.84 &0.8 $^{\pm0.03}$ & 0.51 $^{\pm0.10}$ & 1617  $^{\pm75}$ & 10.07 & 1555  & --434  & 79 & 6~~~\\
NGC 2248 &06 34	35 &26	18	16 &~1.5 &0.16 &1.0 $^{\pm0.04}$ & 0.23 $^{\pm0.05}$ & 1740 $^{\pm80}$ & 10.23 & 1707  & --226  & 250 & 6~~~\\
NGC 2250 &06 32	48 &-05	02	00 &~2.0 &0.29 &0.6 $^{\pm0.02}$ & 0.48 $^{\pm0.10}$ & 1795 $^{\pm83}$ & 10.02 & 1456  & --1031 & --201 & 6~~~\\
NGC 2260 &06 38	03 &-01	28	24 &10.0 &0.55 &0.01 $^{\pm0.00}$ & 1.25 $^{\pm0.20}$ & 1985 $^{\pm90}$ & 10.23 & 1667  & --1071 & --126 & 6~~~\\
NGC 2265 &06 41	41 &11	54	18 &~6.0 &0.60 &0.3 $^{\pm0.01}$ & 0.48 $^{\pm0.10}$ & 2160  $^{\pm100}$& 10.54 & 2010  & --781  & 123 & 6~~~\\
NGC 2312 &06 58	47 &10	17	42 &~3.8 &0.18 &0.33 $^{\pm0.01}$ & 0.16 $^{\pm0.05}$ & 2245$^{\pm103}$& 10.58 & 2030  & --928  & 246 & 6~~~\\
NGC 2318 &06 59	27 &-13	41	54 &10.0 &0.21 &0.05 $^{\pm0.00}$ & 0.65 $^{\pm0.05}$ & 1335 $^{\pm62}$ & 9.48  & 924   & --958  & --104 & 6~~~\\
NGC 2331 &07 06	59 &27	15	42 &~7.0 &0.04 &1.7 $^{\pm0.12}$ & 0.06 $^{\pm0.02}$ & 1285  $^{\pm59}$ & 9.77  & 1222  & --210  & 337 & 6~~~\\
NGC 2338 &07 07	47 &-05	43	12 &~3.5 &0.21 &0.55 $^{\pm0.02}$ & 0.48 $^{\pm0.05}$ & 1800$^{\pm83}$ & 9.95  & 1381  & --1154 & 32 & 6~~~\\
NGC 2348 &07 03	03 &-67	24	42 &~5.0 &0.33 &1.8 $^{\pm0.07}$ & 0.13 $^{\pm0.05}$ & 1070  $^{\pm48}$ & 8.42  & --139 & --969  & --432 & 6~~~\\
NGC 2349 &07 10	48 &-08	35	36 &~9.0 &0.55 &0.75 $^{\pm0.03}$ & 0.61 $^{\pm0.10}$ & 1628 $^{\pm75}$ & 9.76  & 1195  & --1106 & 10 & 6~~~\\
NGC 2351 &07 13	31 &-10	29	12 &~5.0 &0.49 &0.24 $^{\pm0.01}$ & 0.92 $^{\pm0.25}$ & 1882 $^{\pm87}$ & 9.93  & 1336  & --1325 & 2 & 6~~~\\
NGC 2352 &07 13	05 &-24	02	18 &~3.0 &0.90 &0.12 $^{\pm0.00}$ & 0.32 $^{\pm0.15}$ & 1750$^{\pm81}$ & 9.57  & 953   & --1455 & --191 & 6~~~\\
NGC 2364 &07 20	46 &-07	33	00 &~6.5 &0.50 &0.2 $^{\pm0.01}$ & 0.31 $^{\pm0.05}$ & 1919  $^{\pm88}$ & 10.00 & 1401  & --1307 & 101 & 6~~~\\
NGC 2408 &07 40	09 &71	39	18 &14.0 &0.45 &3.0 $^{\pm0.12}$ & 0.03 $^{\pm0.00}$ & 1133  $^{\pm52}$ & 9.44  & 794   & 585    & 557 & 6~~~\\
\end{tabular}}
\end{table*}
\begin{table*}
\centerline{
\begin{tabular}{cccrrrrrrrrrr}
\hline
Cluster  & $\alpha$
            & $\delta$ & R$_{lim.}$ & R$_{c}$& Age  & E$_{B-V}$  & $R_{\odot}$  & $R_\mathrm{gc}$
            & $X_{\odot}$  & $Y_{\odot}$  & $Z$ & Ref.\\
            & $~~^{h}~~^{m}~~^{s}$ & $~~^{\circ}~~{'}~~{''}$ & ${'}$ & ${'}$ & \it Gyr & \it mag  & \it pc  & \it kpc   & \it pc
            & \it pc  & \it pc & \\
\hline
NGC 2455 &07 49	01 &-21	18	06 &~4.0 &0.24 &0.18 $^{\pm0.01}$ & 0.54 $^{\pm0.10}$ & 2650 $^{\pm122}$& 10.14 & 1389  & --2254 & 107 & 6~~~\\
NGC 2459 &07 52	02 &09	33	24 &~2.7 &0.04 &1.6 $^{\pm0.11}$ & 0.03 $^{\pm0.02}$ & 1300 $^{\pm60}$ & 9.64  & 1060  & --640  & 397 & 6~~~\\
NGC 2587 &08 23	25 &-29	30	30 &~2.0 &0.11 &0.1 $^{\pm0.00}$ & 0.23 $^{\pm0.10}$ & 1740 $^{\pm80}$ & 9.26  & 609   & --1624 & 136 & 6~~~\\
NGC 2609 &08 29	32 &-61	06	36 &~2.5 &0.06 &0.8 $^{\pm0.03}$ & 0.23 $^{\pm0.10}$ & 1320 $^{\pm61}$ & 8.46  & --138 & --1280 & --291 & 6~~~\\
NGC 2666 &08 49	47 &44	42	12 &~5.5 &0.30 &3.2 $^{\pm0.13}$ & 0.03 $^{\pm0.01}$ & ~860  $^{\pm40}$ & 9.36  & 664   & 47     & 544 & 6~~~\\
NGC 2678 &08 50	02 &11	20	18 &~6.5 &0.60 &2.3 $^{\pm0.09}$ & 0.03 $^{\pm0.01}$ & ~900  $^{\pm41}$ & 9.24  & 621   & --452  & 469 & 6~~~\\
NGC 2932 &09 35	28 &-46	48	36 &10.5 &0.07 &0.5 $^{\pm0.02}$ & 0.55 $^{\pm0.10}$  & 1525 $^{\pm70}$ & 8.59  & --47  & --1521 & 103 & 6~~~\\
NGC 2995 &09 44	04 &-54	46	48 &~3.5 &0.70 &0.05 $^{\pm0.00}$ & 1.94 $^{\pm0.30}$ & ~380$^{\pm17}$ & 8.46  & --53  & --376  & --8 & 6~~~\\
NGC 3231 &10 26	58 &66	48	55 &~3.5 &0.47 &1.4 $^{\pm0.06}$ & 0.02 $^{\pm0.00}$ & ~715 $^{\pm33}$ & 9.07  & 401   & 314    & 502 & 6~~~\\
NGC 3446 &10 52	12 &-45	08	54 &~7.5 &0.97 &1.0 $^{\pm0.04}$ & 0.16 $^{\pm0.05}$  & 1485 $^{\pm68}$ & 8.32  & --298 & --1417 & 329 & 6~~~\\
NGC 3520 &11 07	08 &-18	01	24 &~1.5 &0.24 &3.2 $^{\pm0.13}$ & 0.03 $^{\pm0.00}$ & 1245 $^{\pm57}$ & 8.57  & --13  & --978  & 771 & 6~~~\\
NGC 3909 &11 49	49 &-48	15	06 &~8.0 &0.99 &2.0 $^{\pm0.08}$ & 0.13 $^{\pm0.05}$  & 1100 $^{\pm50}$ & 8.14  & --409 & --989  & 254 & 6~~~\\
NGC 4230 &12 17	20 &-55	06	06 &~5.0 &0.07 &1.7 $^{\pm0.12}$ & 0.23 $^{\pm0.10}$  & 1445 $^{\pm67}$ & 7.92  & --673 & --1265 & 187 & 6~~~\\
NGC 5155 &13 29	35 &-63	25	30 &~8.5 &0.08 &1.5 $^{\pm0.18}$ & 0.06 $^{\pm0.02}$  & 1070 $^{\pm49}$ & 7.9   & --647 & --852 & --16 & 6~~~\\
NGC 5269 &13 44	44 &-62	54	54 &~1.5 &0.02 &0.16 $^{\pm0.11}$ & 0.52 $^{\pm0.10}$ & 1410$^{\pm65}$ & 7.69  & --886 & --1096 & --16 & 6~~~\\
NGC 5299 &13 50	26 &-59	56	54 &16.5 &0.29 &2.0 $^{\pm0.08}$ & 0.19 $^{\pm0.05}$  & 1111 $^{\pm50}$ & 7.83  & --717 & --847  & 40 & 6~~~\\
NGC 5381 &14 00	41 &-59	35	12 &~5.5 &0.67 &1.6 $^{\pm0.11}$ & 0.06 $^{\pm0.02}$  & 1170 $^{\pm54}$ & 7.77  & --776 & --874 & 43 & 6~~~\\
NGC 5800 &15 01	47 &-51	55	06 &~6.0 &0.14 &0.9 $^{\pm0.04}$ & 0.62 $^{\pm0.10}$  & 2146 $^{\pm99}$ & 6.92  & --1692& --1301 & 222 & 6~~~\\
NGC 5925 &15 27	26 &-54	31	42 &12.0 &0.12 &0.25 $^{\pm0.01}$ & 0.58 $^{\pm0.10}$ & 1040 $^{\pm48}$ & 7.68  & --845 & --606 & 31 & 6~~~\\
NGC 5998 &15 49	34  &-28 35	18 &~4.5 &0.65 &2.2 $^{\pm0.09}$ & 0.16 $^{\pm0.05}$  &  ~981 $^{\pm45}$ & 7.56  & --886 & --257  & 332 & 6~~~\\
NGC 6334 &17 20	49  &-36 06	12 &15.5 &0.99 &0.5 $^{\pm0.02}$ & 1.06 $^{\pm0.25}$  &  1025 $^{\pm47}$ & 7.49  & --1013& --158& 8 & 6~~~\\
NGC 6357 &17 24	43	&-34 12	06 &~2.5 &0.20 &0.4 $^{\pm0.02}$ & 1.35 $^{\pm0.30}$  &  1205 $^{\pm55}$ & 7.3 & --1196& --143 & 19 & 6~~~\\
NGC 6360 &17 24	27	&-29 52	18 &~2.5 &0.74 &0.02 $^{\pm0.00}$ & 1.11 $^{\pm0.20}$ &  1337 $^{\pm62}$ & 7.17 & --1333& --76   & 73 & 6~~~\\
NGC 6374 &17 32	18	&-32 36	00 &~1.8 &0.15 &1.3 $^{\pm0.05}$ & 0.48 $^{\pm0.05}$  &  ~900 $^{\pm41}$ & 7.6  & --897 & --73  & 7 & 6~~~\\
NGC 6421 &17 45	44	&-33 41	36 &~4.0 &0.12 &0.17 $^{\pm0.01}$ & 1.26 $^{\pm0.20}$ &  1505 $^{\pm69}$ & 7.0   & --1500& --107  & --63 & 6~~~\\
NGC 6437 &17 48	24	&-35 21	00 &~7.5 &0.31 &0.2 $^{\pm0.01}$ & 0.71 $^{\pm0.05}$  &  ~943 $^{\pm43}$ & 7.56  & --936 & --91   & --69 & 6~~~\\
NGC 6507 &17 59	50	&-17 27	00 &~6.6 &0.45 &0.40 $^{\pm0.02}$ & 0.85 $^{\pm0.09}$ &  1230 $^{\pm55}$ & 7.3 & --1203 & 246 & 65 & 1~~~\\
NGC 6525 &18 02	06	&11	01	24 &~6.5 &0.88 &2.0 $^{\pm0.08}$ & 0.14 $^{\pm0.03}$  &  1436 $^{\pm66}$ & 7.46  & --1097 & 838  & 393 & 6~~~\\
NGC 6573 &18 13	41	&-22 07	06 &~0.9 &0.15 &0.01 $^{\pm0.00}$ & 2.48 $^{\pm0.20}$ &  ~460 $^{\pm21}$ & 8.05  & --454 & 72     & --17 & 6~~~\\
NGC 6588 &18 20	33	&-63 48	30 &~2.5 &0.13 &1.6 $^{\pm0.11}$ & 0.10 $^{\pm0.03}$  &  ~960 $^{\pm44}$ & 7.68  & --783 & --437  & --342 & 6~~~\\
NGC 6595 &18 17	04	&-19 51	54 &~2.0 &0.27 &0.45 $^{\pm0.02}$ & 0.94 $^{\pm0.10}$ &  1640 $^{\pm76}$ & 6.90 & --1607 & 325 & --49 & 6~~~\\
NGC 6605 &18 18	21	&-14 56	42 &~8.5 &0.46 &0.6 $^{\pm0.02}$ & 0.52 $^{\pm0.10}$  &  ~889 $^{\pm40}$ & 7.65  & --855 & 244 & 5 & 6~~~\\
NGC 6625 &18 22	50	&-11 57	42 &~7.7 &0.22 &0.50 $^{\pm0.03}$ & 1.21 $^{\pm0.13}$ &  1335 $^{\pm60}$ & 7.25 & --1262 & 435 & 19 & 1~~~\\
NGC 6645 &18 32	37	&-16 53	00 &~7.4 &0.79 &0.40 $^{\pm0.03}$ & 0.36 $^{\pm0.07}$ &  1245 $^{\pm55}$ & 7.31 & --1195 & 338 & --82 & 1~~~\\
NGC 6647 &18 32	50	&-17 13	56 &~6.5 &0.75 &1.60 $^{\pm0.05}$ & 0.54 $^{\pm0.10}$ &  2200 $^{\pm100}$ & 6.4 & --2119 & 577 & --137 & 1~~~\\
NGC 6659 &18 33	59	&23	35	42 &~7.0 &0.11 &4.0 $^{\pm0.16}$ & 0.10 $^{\pm0.03}$  &  1155 $^{\pm53}$ & 7.85  & --682 & 888    & 282 & 6~~~\\
NGC 6698 &18 48	04	&-25 52	42 &~5.5 &0.40 &1.9 $^{\pm0.07}$ & 0.32 $^{\pm0.05}$  &  1150 $^{\pm53}$ & 7.37  & --1115& 182    & --215 & 6~~~\\
NGC 6724 &18 56	46	&10	25	42 &~3.0 &0.13 &0.9 $^{\pm0.03}$ & 1.00 $^{\pm0.10}$  &  1105 $^{\pm51}$ & 7.73  & --809 & 750    & 69 & 6~~~\\
NGC 6735 &19 00	37	&00	28	30 &~6.0 &0.03 &0.5 $^{\pm0.02}$ & 0.87 $^{\pm0.15}$  &  1466 $^{\pm68}$ & 7.34 & --1209& 828 & --47 & 6~~~\\
NGC 6737 &19 02	20	&-18 32	59 &~4.4 &0.41 &0.50 $^{\pm0.02}$ & 0.76 $^{\pm0.11}$ &  2120 $^{\pm95}$ & 6.51 & --1988 & 624 & --392 & 1~~~\\
NGC 6743 &19 01	20	&29	16	36 &~3.5 &0.08 &1.4 $^{\pm0.05}$ & 0.19 $^{\pm0.05}$  &  1111 $^{\pm51}$ & 8.01  & --539 & 948    & 211 & 6~~~\\
NGC 6773 &19 15	03	&04	52	54 &~4.3 &0.12 &0.1 $^{\pm0.00}$ & 1.16 $^{\pm0.20}$  &  2160 $^{\pm100}$ & 6.98  & --1653& 1386   & --113 & 6~~~\\
NGC 6775 &19 16	48	&-00 55	24 &~1.2 &0.07 &0.9 $^{\pm0.03}$ & 0.48 $^{\pm0.05}$  &  1185 $^{\pm55}$ & 7.58  & --947 & 705    & --108 & 6~~~\\
NGC 6795 &19 26	22	&03	30	54 &~4.0 &0.78 &0.95 $^{\pm0.04}$ & 0.45 $^{\pm0.05}$ &  1320 $^{\pm61}$ & 7.54  & --1004& 845    & --141 & 6~~~\\
NGC 6815 &19 40	44	&26	45	30 &15.0 &0.10 &0.15 $^{\pm0.01}$ & 1.10 $^{\pm0.20}$ &  2024 $^{\pm93}$ & 7.76 & --945 & 1788 & 72 & 6~~~\\
NGC 6832 &19 48	15	&59	25	18 &12.0 &0.11 &3.0 $^{\pm0.12}$ & 0.10 $^{\pm0.02}$  &  1750 $^{\pm81}$ & 8.74  & 59    & 1678    & 493 & 6~~~\\
NGC 6837 &19 53	08	&11	41	54 &~2.3 &0.94 &1.0 $^{\pm0.04}$ & 0.25 $^{\pm0.02}$  &  ~943 $^{\pm43}$ & 7.93  & --594 & 721    & --131 & 6~~~\\
\end{tabular}}
\end{table*}
\begin{table*}
\centerline{
\begin{tabular}{cccrrrrrrrrrr}
\hline
Cluster  & $\alpha$
            & $\delta$ & R$_{lim.}$ & R$_{c}$& Age  & E$_{B-V}$  & $R_{\odot}$  & $R_\mathrm{gc}$
            & $X_{\odot}$  & $Y_{\odot}$  & $Z$ & Ref.\\
            & $~~^{h}~~^{m}~~^{s}$ & $~~^{\circ}~~{'}~~{''}$ & ${'}$ & ${'}$ & \it Gyr & \it mag  & \it pc  & \it kpc   & \it pc
            & \it pc  & \it pc & \\
\hline
NGC 6839 &19 54	33	&17	56	18 &~3.0 &0.90 &1.4 $^{\pm0.05}$ & 0.29 $^{\pm0.05}$  &  1410 $^{\pm65}$ & 7.8   & --783 & 1166   & --127 & 6~~~\\
NGC 6840 &19 55	18	&12	07	36 &~3.0 &0.04 &1.3 $^{\pm0.04}$ & 0.25 $^{\pm0.05}$  &  1970 $^{\pm90}$ & 7.42  & --1223& 1518   & --283 & 6~~~\\
NGC 6843 &19 56	06	&12	09	48 &~2.5 &0.26 &1.3 $^{\pm0.04}$  & 0.30 $^{\pm0.05}$ &  1945 $^{\pm90}$ & 7.44  & --1203& 1501   & --284 & 6~~~\\
NGC 6846 &19 56	28	&32	20	54 &~2.4 &0.12 &0.55 $^{\pm0.02}$ & 0.68 $^{\pm0.05}$ &  1445 $^{\pm67}$ & 8.09 & --525 & 1345 & 48 & 6~~~\\
NGC 6847 &19 56	37	&30	12	48 &10.0 &0.74 &0.5 $^{\pm0.02}$ & 0.58 $^{\pm0.05}$  &  1894 $^{\pm87}$ & 7.95 & --743 & 1742 & 26 & 6~~~\\
NGC 6856 &19 59	17	&56	07	48 &~1.6 &0.08 &1.8 $^{\pm0.06}$ & 0.16 $^{\pm0.02}$  &  1704 $^{\pm79}$ & 8.66  & --9   & 1657   & 398 & 6~~~\\
NGC 6858 &20 02	56	&11	15	30 &~5.0 &0.59 &2.5 $^{\pm0.10}$ & 0.13 $^{\pm0.02}$  &  1310 $^{\pm60}$ & 7.75  & --805 & 1007   & --234 & 6~~~\\
NGC 6859 &20 03	49	&00	26	36 &~5.0 &0.07 &3.0 $^{\pm0.12}$ & 0.19 $^{\pm0.05}$  &  1335 $^{\pm62}$ & 7.56  & --957 & 856    & --364 & 6~~~\\
NGC 6873 &20 07	13	&21	06	06 &~7.5 &0.90 &0.88 $^{\pm0.04}$ & 0.35 $^{\pm0.05}$ &  1250 $^{\pm58}$ & 7.96  & --613 & 1081   & --134 & 6~~~\\
NGC 6895 &20 16	29	&50	13	48 &~8.0 &0.85 &1.0 $^{\pm0.04}$ & 0.35 $^{\pm0.05}$  &  1141 $^{\pm53}$ & 8.5   & --81  & 1126   & 164 & 6~~~\\
NGC 6904 &20 21	48	&25	44	24 &~4.0 &0.69 &1.0 $^{\pm0.04}$ & 0.39 $^{\pm0.05}$  &  1355 $^{\pm62}$ & 8.05  & --545 & 1232  & --149 & 6~~~\\
NGC 6938 &20 34	42	&22	12	54 &~3.6 &0.29 &1.3 $^{\pm0.04}$ & 0.13 $^{\pm0.05}$  &  1250 $^{\pm58}$ & 8.05  & --521 & 1112  & --233 & 6~~~\\
NGC 6950 &20 41	04	&16	37	06 &~7.5 &0.41 &1.8 $^{\pm0.05}$ & 0.06 $^{\pm0.02}$  &  1070 $^{\pm49}$ & 8.04  & --499 & 904   & --281 & 6~~~\\
NGC 7005 &21 01	57	&-12 52	50 &~2.0 &0.22 &2.5 $^{\pm0.10}$ & 0.03 $^{\pm0.01}$  &  1033 $^{\pm48}$ & 7.69  & --689 & 497    & --588 & 6~~~\\
NGC 7011 &21 01	49	&47	21	12 &~2.2 &0.14 &0.4 $^{\pm0.01}$ & 1.08 $^{\pm0.10}$  &  1236 $^{\pm57}$ & 8.55  & --40  & 1235   & 13 & 6~~~\\
NGC 7023 &21 01	35	&68	10	12 &~7.2 &0.33 &0.12 $^{\pm0.00}$ & 1.10 $^{\pm0.10}$ &  ~560 $^{\pm26}$ & 8.65  & 132 & 527  & 137 & 6~~~\\
NGC 7024 &21 06	09	&41	29	18 &~2.5 &0.23 &0.5 $^{\pm0.02}$ & 1.10 $^{\pm0.10}$  &  1760 $^{\pm81}$ & 8.51  & --175 & 1747   & --119 & 6~~~\\
NGC 7037 &21 10	54	&33	45	48 &~3.0 &0.01 &2.1 $^{\pm0.08}$ & 0.16 $^{\pm0.05}$  &  1485 $^{\pm68}$ & 8.35  & --276 & 1437   & --252 & 6~~~\\
NGC 7050 &21 15	12	&36	10	24 &~3.5 &0.45 &2.0 $^{\pm0.08}$ & 0.16 $^{\pm0.05}$  &  1179 $^{\pm54}$ & 8.41  & --171 & 1152   & --180 & 6~~~\\
NGC 7055 &21 19	30	&57	34	12 &~2.5 &0.08 &0.8 $^{\pm0.03}$ & 1.10 $^{\pm0.10}$  &  1275 $^{\pm59}$ & 8.76  & 165   & 1258   & 124 & 6~~~\\
NGC 7071 &21 26	39	&47	55	12 &~4.0 &0.15 &0.3 $^{\pm0.01}$ & 1.14 $^{\pm0.20}$  &  1684 $^{\pm78}$ & 8.71  & 42    & 1682   & --59 & 6~~~\\
NGC 7084 &21 32	33	&17	30	30 &~8.0 &0.90 &1.5 $^{\pm0.06}$ & 0.10 $^{\pm0.05}$  &  ~765 $^{\pm35}$ & 8.27  & --239 & 655    & --315 & 6~~~\\
NGC 7093 &21 34	21	&45	57	54 &~6.5 &0.04 &0.9 $^{\pm0.04}$ & 0.61 $^{\pm0.05}$  &  1785 $^{\pm82}$ & 8.72  & 32    & 1780   & --135 & 6~~~\\
NGC 7127 &21 43	41	&54	37	48 &~2.5 &0.15 &0.4 $^{\pm0.02}$ & 0.90 $^{\pm0.05}$  &  1445 $^{\pm67}$ & 8.82  & 199 & 1431 & 29 & 6~~~\\
NGC 7129 &21 42	59	&66	06	48 &~3.5 &0.21 &0.12 $^{\pm0.01}$ & 0.97 $^{\pm0.05}$ &  1070 $^{\pm49}$ & 8.84 & 280 & 1016 & 184 & 6~~~\\
NGC 7134 &21 48	55	&-12 58	24 &~1.5 &0.10 &3.3 $^{\pm0.13}$ & 0.06 $^{\pm0.02}$  &  1065 $^{\pm49}$ & 7.74  & --558 & 502    & --755 & 6~~~\\
NGC 7175 &21 58	46	&54	49	06 &16.0 &0.18 &0.25 $^{\pm0.01}$ & 0.87 $^{\pm0.05}$ &  1930 $^{\pm89}$ & 9.03 & 326 & 1902 & --3 & 6~~~\\
NGC 7193 &22 03	03	&10	48	06 &~6.5 &0.39 &4.5 $^{\pm0.18}$ & 0.03 $^{\pm0.00}$  &  1080 $^{\pm50}$ & 8.2 & --304 & 839    & --608 & 6~~~\\
NGC 7352 &22 39	43	&57	23	42 &~4.5 &0.03 &0.05 $^{\pm0.00}$ & 1.10 $^{\pm0.20}$ &  2550 $^{\pm117}$ & 9.52  & 698   & 2452   & --47 & 6~~~\\
NGC 7394 &22 50	23	&52	08	06 &~4.5 &0.20 &0.6 $^{\pm0.02}$ & 0.35 $^{\pm0.05}$  &  1310 $^{\pm60}$ & 8.92  & 332   & 1259   & --147 & 6~~~\\
NGC 7429 &22 56	00	&59	58	24 &~7.0 &0.06 &0.04 $^{\pm0.00}$ & 1.16 $^{\pm0.10}$ &  1190 $^{\pm55}$ & 8.96  & 387 & 1125 & 6 & 6~~~\\
NGC 7686 &23 30	07	&49	08	00 &~8.0 &0.80 &2.0 $^{\pm0.08}$ & 0.20 $^{\pm0.05}$  &  1534 $^{\pm71}$ & 9.13  & 502   & 1416   & --307 & 6~~~\\
NGC 7708 &23 35	01	&72	50	00 &12.0 &0.90 &2.0 $^{\pm0.08}$ & 0.42 $^{\pm0.05}$  &  1607 $^{\pm74}$ & 9.35  & 726   & 1401   & 301 & 6~~~\\
NGC 7795 &23 58	37	&60	02	06 &10.5 &0.19 &0.45 $^{\pm0.02}$ & 1.00 $^{\pm0.10}$ &  2105 $^{\pm97}$ & 9.62  & 935   & 1885   & --79 & 6~~~\\
NGC 7801 &00 00	21	&50	44	30 &~4.0 &0.98 &1.7 $^{\pm0.12}$ & 0.17 $^{\pm0.05}$  &  1275 $^{\pm60}$ & 9.11  & 523   & 1136   & --250 & 6~~~\\
NGC 7826 &00 05	17	&-20 41	30 &10.0 &0.80 &2.2 $^{\pm0.09}$ & 0.03 $^{\pm0.01}$  &  ~620 $^{\pm29}$ & 8.23  & --62  & 117    & --606 & 6~~~\\
NGC 7833 &00 06	31	&27	38	30 &~1.3 &0.10 &2.0 $^{\pm0.08}$ & 0.06 $^{\pm0.02}$  &  1410 $^{\pm65}$ & 9.10  & 416   & 1090   & --792 & 6~~~\\
Berkeley ~~1 &00 09	36	&60	28	30 &2.5 &0.14 &0.4~$^{\pm0.03}$  & 0.78 $^{\pm0.06}$ &  2420 $^{\pm110}$ & 9.9~ & 1128 & 2139 & --84 & 2~~~\\
Berkeley ~~6 &01 51	11	&61	03	40 &3.0 &0.15 &0.1~$^{\pm0.00}$  & 0.78 $^{\pm0.05}$ &  2300 $^{\pm105}$ & 10.1 & 1481 & 1759 & --38 & 2~~~\\
Berkeley ~26 &06 50	18	&05	45	00 &2.6 &0.13 &0.6~$^{\pm0.03}$  & 0.54 $^{\pm0.05}$ &  2720 $^{\pm120}$ & 11.0 & 2407 & --1262 & 112 & 2~~~\\
Berkeley ~37 &07 20	24	&-01 06	00 &3.5 &0.69 &0.9~$^{\pm0.04}$  & 0.12 $^{\pm0.02}$ &  4555 $^{\pm210}$ & 12.4 & 3605 & --2740 & 470 & 2~~~\\
Berkeley ~43 &19 15	36	&11	13	00 &4.5 &0.57 &0.4~$^{\pm0.03}$  & 1.52 $^{\pm0.14}$ &  1355 $^{\pm60}$ & 7.6~ & --947 & 969 & --4 & 2~~~\\
Berkeley ~45 &19 19	12	&15	43	00 &3.5 &0.14 &0.6~$^{\pm0.03}$  & 0.82 $^{\pm0.05}$ &  2300 $^{\pm105}$ & 7.2~ & --1477 & 1763 & 46 & 2~~~\\
Berkeley ~47 &19 28	36	&17	22	06 &2.0 &0.03 &0.16 $^{\pm0.01}$ & 1.06 $^{\pm0.10}$ &  1420 $^{\pm65}$ & 7.7~ & --863 & 1127 & --1.4 & 2~~~\\
Berkeley ~49 &19 59	31	&34	38	48 &2.4 &0.17 &0.16 $^{\pm0.01}$ & 1.57 $^{\pm0.30}$ &  2035 $^{\pm110}$ & 8.1~ & --662 & 1922 & 91 & 2~~~\\
Berkeley ~50 &20 10	24	&34	58	00 &3.5 &0.12 &0.25 $^{\pm0.01}$ & 0.97 $^{\pm0.07}$ &  2100 $^{\pm100}$ & 8.1~ & --633 & 2002 & 31 & 2~~~\\
Berkeley ~51 &20 11	54	&34	24	06 &1.5 &0.13 &0.15 $^{\pm0.02}$ & 1.66 $^{\pm0.14}$ &  3200 $^{\pm145}$ & 8.1~ & --981 & 3046 & 16 & 2~~~\\
\end{tabular}}
\end{table*}
\begin{table*}
\centerline{
\begin{tabular}{cccrrrrrrrrrr}
\hline
Cluster  & $\alpha$
            & $\delta$ & R$_{lim.}$ & R$_{c}$& Age  & E$_{B-V}$  & $R_{\odot}$  & $R_\mathrm{gc}$
            & $X_{\odot}$  & $Y_{\odot}$  & $Z$ & Ref.\\
            & $~~^{h}~~^{m}~~^{s}$ & $~~^{\circ}~~{'}~~{''}$ & ${'}$ & ${'}$ & \it Gyr & \it mag  & \it pc  & \it kpc   & \it pc
            & \it pc  & \it pc & \\
\hline
Berkeley ~61 &00 48	30	&67	12	00 &3.5 &0.19 &0.8~$^{\pm0.05}$  & 1.09 $^{\pm0.11}$ &  3335 $^{\pm150}$ & 10.7 & 1794 & 2800 & 252 & 2~~~\\
Berkeley ~63 &02 19	36	&63	43	00 &3.6 &0.35 &0.5~$^{\pm0.02}$  & 0.90 $^{\pm0.09}$ &  3305 $^{\pm150}$ & 11.0 & 2231 & 2434 & 144 & 2~~~\\
Berkeley ~72 &05 50	18	&22	12	00 &3.5 &0.08 &0.6~$^{\pm0.02}$  & 0.79 $^{\pm0.07}$ &  3810 $^{\pm175}$ & 12.3 & 3783 & --419 & --171 & 2~~~\\
Berkeley ~76 &07 06	40	&-11 44	00 &4.5 &0.96 &0.8~$^{\pm0.04}$  & 0.73 $^{\pm0.05}$ &  2505 $^{\pm115}$ & 10.4 & 1764 & --1770 & --87 & 2~~~\\
Berkeley ~84 &20 04	43	&33	54	18 &1.1 &0.06 &0.12 $^{\pm0.01}$ & 0.76 $^{\pm0.05}$ &  2025 $^{\pm95}$ & 8.1~ & --661 & 1912 & 45 & 2~~~\\
Berkeley ~89 &20 24	36	&46	03	00 &2.5 &0.08 &0.85~$^{\pm0.05}$ & 1.03 $^{\pm0.10}$ &  3005 $^{\pm135}$ & 8.7~ & --357 & 2973 & 253 & 2~~~\\
Berkeley ~90 &20 35	18	&46	50	00 &2.5 &0.22 &0.1~$^{\pm0.01}$  & 1.15 $^{\pm0.10}$ &  2430 $^{\pm70}$ & 8.6~ & --216 & 2415 & 160 & 2~~~\\
Berkeley ~91 &21 10	52	&48	32	12 &1.7 &0.13 &0.5~$^{\pm0.02}$  & 1.00 $^{\pm0.09}$ &  2400 $^{\pm110}$ & 8.8~ & 2.7 & 2400 & 5.5 & 2~~~\\
Berkeley ~95 &22 28	18	&59	08	00 &2.4 &0.29 &0.15 $^{\pm0.02}$ & 1.21 $^{\pm0.12}$ &  1900 $^{\pm85}$ & 9.2~ & 507 & 1830 & 40 & 2~~~\\
Berkeley ~97 &22 39	30	&59	01	00 &2.0 &0.10 &0.02 $^{\pm0.00}$ & 0.75 $^{\pm0.05}$ &  1800 $^{\pm85}$ & 9.2~ & 516 & 1724 & 12 & 2~~~\\
Berkeley 100 &23 25	58	&63	46	48 &2.2 &0.29 &0.16~$^{\pm0.01}$ & 1.21 $^{\pm0.11}$ &  3355 $^{\pm155}$ & 10.3 & 1345 & 3070 & 144 & 2~~~\\
Berkeley 101 &23 32	47	&64	12	30 &2.2 &0.23 &0.7~$^{\pm0.05}$  & 1.11 $^{\pm0.10}$ &  2500 $^{\pm115}$ & 9.8~ & 1036 & 2272 & 115 & 2~~~\\
Berkeley 102 &23 38	42	&56	38	00 &3.3 &0.28 &0.6~$^{\pm0.03}$  & 0.69 $^{\pm0.05}$ &  2600 $^{\pm120}$ & 9.8~ & 1013 & 2384 & --219 & 2~~~\\
Berkeley 103 &23 45	12	&59	18	00 &2.5 &0.36 &0.5~$^{\pm0.02}$  & 1.00 $^{\pm0.11}$ &  2100 $^{\pm95}$ & 9.6~ & 872 & 1908 & --91 & 2~~~\\
Kronberger ~~2 &18 21 19 &-14 17 12&2.5 &0.08 &0.10 $^{\pm0.00}$ & 1.10 $^{\pm0.10}$ &  3065 $^{\pm140}$ & 5.6 & --2934 & 887 & ~~1.5 & 4~~~\\
Kronberger ~~3 &19 39 00 &06 46	00 &1.0 &0.03 &1.60 $^{\pm0.11}$ & 0.38 $^{\pm0.08}$ &  1870 $^{\pm85}$ & 7.3 & --1324 & 1299 & --240 & 4~~~\\
Kronberger ~~5 &19 46 05 &27 50	00 &3.5 &0.08 &0.16 $^{\pm0.02}$ & 2.10 $^{\pm0.40}$ &  615~ $^{\pm30}$ & 8.2 & --273 & 551 & 17 & 4~~~\\
Kronberger 12  &06 14 16 &22 29	52 &1.4 &0.17 &0.16 $^{\pm0.02}$ & 0.97 $^{\pm0.05}$ &  1775 $^{\pm80}$ & 10.3 & 1753 & --271 & 74 & 4~~~\\
Kronberger 13  &19 25 15 &13 56	44 &1.2 &0.08 &0.40 $^{\pm0.01}$ & 1.13 $^{\pm0.11}$ &  1380 $^{\pm65}$ & 7.7 & --902 & 1044 & --24 & 4~~~\\
Kronberger 18  &05 18 36 &37 37	18 &4.0 &0.36 &0.10 $^{\pm0.00}$ & 1.29 $^{\pm0.12}$ &  3250 $^{\pm150}$ & 11.7 & 3197 & 584 & 1.9 & 4~~~\\
Kronberger 23  &23 05 59 &60 15	14 &0.9 &0.15 &0.10 $^{\pm0.01}$ & 1.35 $^{\pm0.12}$ &  1740 $^{\pm80}$ & 9.2 & 601 & 1633 & 0.44 & 4~~~\\
Kronberger 25  &18 22 40 &-14 43 41&0.8 &0.10 &0.05 $^{\pm0.00}$ & 1.32 $^{\pm0.11}$ &  1220 $^{\pm55}$ & 7.3 & --1169 & 348 & --10 & 4~~~\\
Kronberger 28  &20 06 32 &35 34	34 &0.75 &0.09 &0.40 $^{\pm0.01}$ & 2.26 $^{\pm0.42}$ &  550~ $^{\pm25}$ & 8.4 & --165 & 524 & 18 & 4~~~\\
Kronberger 52  &19 58 08 &30 53	18 &1.2 &0.06 &0.13 $^{\pm0.01}$ & 2.45 $^{\pm0.44}$ &  705~ $^{\pm30}$ & 8.3 & --268 & 652 & 10.5 & 4~~~\\
Kronberger 54  &20 03 08 &31 58	01 &0.8 &0.11 &0.25 $^{\pm0.02}$ & 0.94 $^{\pm0.04}$ &  1715 $^{\pm80}$ & 8.0 & --612 & 1602 & 15.5 & 4~~~\\
Kronberger 55  &23 53 09 &62 47	12 &1.1 &0.07 &0.40 $^{\pm0.02}$ & 1.13 $^{\pm0.11}$ &  1260 $^{\pm60}$ & 9.1 & 559 & 1129 & 15 & 4~~~\\
Kronberger 57  &20 23 58 &36 36	17 &2.2 &0.16 &0.16 $^{\pm0.02}$ & 1.06 $^{\pm0.09}$ &  1295 $^{\pm60}$ & 8.3 & --328 & 1253 & --11 & 4~~~\\
Kronberger 58  &20 20 48 &41 12	17 &0.25 &0.01&0.16 $^{\pm0.02}$ & 1.35 $^{\pm0.12}$ &  1515 $^{\pm70}$ & 8.3 & --295 & 1484 & 69.5 & 4~~~\\
Kronberger 59  &20 23 50 &40 08	53 &1.0 &0.04 &0.10 $^{\pm0.00}$ & 0.84 $^{\pm0.11}$ &  780~ $^{\pm35}$ & 8.4 & --144 & 766 & 0.05 & 4~~~\\
Kronberger 60  &06 04 10 &31 29	44 &1.6 &0.12 &0.80 $^{\pm0.03}$ & 0.84 $^{\pm0.11}$ &  1960 $^{\pm90}$ & 10.5 & 1953 & 6.6 & 162 & 4~~~\\
Kronberger 68  &20 00 36 &30 35	23 &2.2 &0.18 &0.20 $^{\pm0.03}$ & 2.19 $^{\pm0.40}$ &  710~ $^{\pm30}$ & 8.2 & --270 & 567 & 3.1 & 4~~~\\
Kronberger 72  &20 12 19 &37 53	27 &2.0 &0.23 &0.50 $^{\pm0.02}$ & 0.55 $^{\pm0.08}$ &  1055 $^{\pm50}$ & 8.3 & --271 & 1019 & 39 & 4~~~\\
Kronberger 73  &20 13 47 &36 44	55 &1.2 &0.09 &0.40 $^{\pm0.02}$ & 0.97 $^{\pm0.05}$ &  1695 $^{\pm80}$ & 8.2 & --458 & 1632 & 37 & 4~~~\\
Kronberger 74  &20 17 57 &36 45	37 &1.1 &0.05 &1.00 $^{\pm0.04}$ & 0.87 $^{\pm0.05}$ &  1760 $^{\pm80}$ & 8.2 & --462 & 1698 & 18 & 4~~~\\
Kronberger 80  &21 11 50 &52 22	48 &1.7 &0.16 &0.70 $^{\pm0.03}$ & 1.29 $^{\pm0.10}$ &  1355 $^{\pm60}$ & 8.7 & 69 & 1352 & 66 & 4~~~\\
Kronberger 84  &21 35 32 &53 30	49 &2.2 &0.19 &0.60 $^{\pm0.02}$ & 0.61 $^{\pm0.07}$ &  1075 $^{\pm50}$ & 8.7 & 117 & 1068 & 21 & 4~~~\\
Kronberger 85  &07 58 21 &-34 46 11&1.5 &0.11 &0.30 $^{\pm0.01}$ & 1.00 $^{\pm0.10}$ &  1525 $^{\pm70}$ & 9.1 & 497 & --1440 & --76 & 4~~~\\
Czernik ~1 &00 07 38 &61 28	30 &2.5 &0.12 &0.005 $^{\pm0.00}$ & 1.23 $^{\pm0.11}$ &  2530 $^{\pm115}$  & 9.9 & 1177& 2239 & --42 & 4~~~\\
Czernik ~2 &00 43 42 &60 09	00 &5.8 &0.08 &0.10 $^{\pm0.01}$ & 0.74  $^{\pm0.12}$ &  1775 $^{\pm80}$ & 9.6 & 939 & 1504 & --84 & 4~~~\\
Czernik ~3 &01 03 06 &62 47	00 &2.4 &0.14 &0.10 $^{\pm0.01}$ & 1.42  $^{\pm0.14}$ &  1410 $^{\pm65}$ & 9.4 & 794 & 1165 & ~--1.4 & 4~~~\\
Czernik ~4 &01 35 24 &61 26	00 &2.6 &0.77 &0.10 $^{\pm0.01}$ & 1.06  $^{\pm0.10}$ &  1630 $^{\pm75}$ & 9.6 & 1007& 1281 & --28 & 4~~~\\
Czernik ~5 &01 55 06 &61 20	00 &1.5 &0.15 &0.70 $^{\pm0.04}$ & 1.23  $^{\pm0.10}$ &  2205 $^{\pm100}$ & 10.1& 1432& 1677 & --23 & 4~~~\\
Czernik ~6 &02 02 00 &62 50	00 &2.6 &0.66 &0.25 $^{\pm0.06}$ & 1.00  $^{\pm0.10}$ &  2530 $^{\pm115}$ & 10.3& 1656& 1912 & 47 & 4~~~\\
Czernik ~7 &02 02 24 &62 15	00 &2.5 &0.31 &0.10 $^{\pm0.01}$ & 1.19  $^{\pm0.11}$ &  2235 $^{\pm100}$ & 10.1& 1469& 1684 & 20 & 4~~~\\
Czernik 10 &02 33 54 &60 10	00 &2.2 &0.06 &0.20 $^{\pm0.02}$ & 1.71 $^{\pm0.13}$ &  1575 $^{\pm70}$ & 9.7 & 1120& 1107 & ~--6 & 4~~~\\
Czernik 11 &02 36 35 &59 38	00 &3.0 &0.26 &0.35 $^{\pm0.03}$ & 1.00 $^{\pm0.10}$ &  1210 $^{\pm55}$ & 9.4 & 868 & ~842 & --12 & 4~~~\\
Czernik 12 &02 39 12 &54 55	00 &3.0 &0.85 &0.40 $^{\pm0.02}$ & 0.45 $^{\pm0.07}$ &  2090 $^{\pm95}$ & 10.2& 1550& 1392 & --173 & 4~~~\\
\end{tabular}}
\end{table*}
\begin{table*}
\centerline{
\begin{tabular}{cccrrrrrrrrrr}
\hline
Cluster  & $\alpha$
            & $\delta$ & R$_{lim.}$ & R$_{c}$& Age  & E$_{B-V}$  & $R_{\odot}$  & $R_\mathrm{gc}$
            & $X_{\odot}$  & $Y_{\odot}$  & $Z$ & Ref.\\
            & $~~^{h}~~^{m}~~^{s}$ & $~~^{\circ}~~{'}~~{''}$ & ${'}$ & ${'}$ & \it Gyr & \it mag  & \it pc  & \it kpc   & \it pc
            & \it pc  & \it pc & \\
\hline
Czernik 14 &03 16 54 &58 36	00 &2.0 &0.03 &0.25 $^{\pm0.07}$ & 1.71 $^{\pm0.13}$ & 2175 $^{\pm100}$ & 10.3& 1688& 1371 & 35 & 4~~~\\
Czernik 15 &03 23 12 &52 15	00 &2.4 &0.21 &0.02 $^{\pm0.00}$ & 1.00 $^{\pm0.10}$ &  1155 $^{\pm55}$ & 9.5 & 945 & ~659 & --80 & 4~~~\\
Czernik 16 &03 30 48 &52 39	00 &5.0 &0.55 &0.20 $^{\pm0.03}$ & 1.29 $^{\pm0.10}$ &  2580 $^{\pm120}$ & 10.7& 2132& 1447 & --134 & 4~~~\\
Czernik 17 &03 52 24 &61 57	00 &3.6 &0.39 &0.40 $^{\pm0.04}$ & 0.87 $^{\pm0.05}$ &  2120 $^{\pm95}$ & 10.3& 1673& 1281 & 228 & 4~~~\\
Czernik 25 &06 13 06 &06 59	00 &5.0 &0.60 &0.13 $^{\pm0.02}$ & 0.58 $^{\pm0.07}$ &  1980 $^{\pm90}$ & 10.4& 1824& --748& --181 & 4~~~\\
Czernik 26 &06 30 48 &-04 13 00 &2.7&0.08 &0.17 $^{\pm0.02}$ & 0.45 $^{\pm0.06}$ &  1200 $^{\pm55}$ & 9.5 & 984 & --673& --136 & 4~~~\\
Czernik 30 &07 31 18 &-09 58 00 &3.6 &0.32 &0.13 $^{\pm0.01}$ & 0.35 $^{\pm0.04}$ &  2275 $^{\pm105}$ & 10.2& 1566& --1642& 165 & 4~~~\\
Czernik 37 &17 53 17 &-27 22 10 &1.8 &0.10 &0.60 $^{\pm0.03}$ & 1.03 $^{\pm0.10}$ &  1730 $^{\pm80}$ & 6.77 & --1728 & 67 & --19 & 1~~~\\
Czernik 38 &18 49 42 &04 56	00 &5.0 &0.11 &0.01 $^{\pm0.00}$ & 2.03 $^{\pm0.36}$ &  1910 $^{\pm90}$ & 7.1 & --1521& 1152& 88 & 4~~~\\
Czernik 39 &19 07 44 &04 20	00 &2.5 &0.04 &0.01 $^{\pm0.00}$ & 2.90 $^{\pm0.45}$ &  1340 $^{\pm60}$ & 7.5 & --1046& ~836& --38 & 4~~~\\
Czernik 42 &22 39 48 &59 54	54 &3.5 &0.46 &0.01 $^{\pm0.00}$ & 1.74 $^{\pm0.12}$ &  2585 $^{\pm120}$ & 9.6 & 761 & 2470 & 52 & 4~~~\\
Czernik 44 &23 33 30 &61 57	00 &4.0 &0.08 &0.16 $^{\pm0.04}$ & 1.48 $^{\pm0.10}$ &  3450 $^{\pm160}$ & 10.4& 1398& 3154 & 27 & 4~~~\\
Czernik 45 &23 56 18 &64 33	00 &2.5 &0.19 &0.02 $^{\pm0.00}$ & 1.45 $^{\pm0.11}$ &  2530 $^{\pm115}$ & 9.9 & 1150& 2251 & 102 & 4~~~\\
Dol-Dzim ~1 &02	47 30 &17 16 00 &7.0 &0.16 &5.00 $^{\pm0.25}$ & 0.10 $^{\pm0.03}$ &  960~ $^{\pm45}$ & 9.4 & 710 & ~278 & --584 & 4~~~\\
Dol-Dzim ~2 &05	23 54 &11 28 00 &5.0 &0.38 &0.80 $^{\pm0.03}$ & 0.65 $^{\pm0.11}$ &  1220 $^{\pm55}$ & 9.7 & 1159 & --251 & --286 & 4~~~\\
Dol-Dzim ~3 &05	33 42 &26 29 00 &4.5 &0.19 &0.40 $^{\pm0.02}$ & 0.77 $^{\pm0.13}$ &  2530 $^{\pm115}$ & 11.0& 2525 & ~--30 & --156 & 4~~~\\
Dol-Dzim ~4 &05	35 54 &25 57 00 &12.0 &0.26 &0.10 $^{\pm0.00}$ & 1.10 $^{\pm0.18}$ &  1220 $^{\pm55}$ & 9.7 & 1217 & ~--30 & --73 & 4~~~\\
Dol-Dzim ~5 &16	27 24 &38 04 00 &15.0 &0.55 &2.0  $^{\pm0.09}$ & 0.02 $^{\pm0.00}$ &  900~ $^{\pm40}$ & 8.1 & --316~ & 567  & 624~ & 4~~~\\
Dol-Dzim ~6 &16	45 24 &38 21 00 &4.0  &0.53 &5.0  $^{\pm0.26}$ & 0.03 $^{\pm0.00}$ &  820~ $^{\pm35}$ & 8.1 & --297~ & 549  & 531~ & 4~~~\\
Dol-Dzim ~7 &17	10 36 &15 32 00 &3.0  &0.88 &2.0  $^{\pm0.11}$ & 0.13 $^{\pm0.02}$ &  1140 $^{\pm50}$ & 7.6 & --802~ & 589  & 555~ & 4~~~\\
Dol-Dzim ~8 &17	26 12 &24 11 00 &8.0 &0.90 &3.0  $^{\pm0.12}$ & 0.06 $^{\pm0.00}$ &  2330 $^{\pm100}$ & 7.1 & --1392 & 1494 & 1123 & 4~~~\\
Dol-Dzim ~9 &18	08 48 &31 32 00 &12.0 &0.30 &2.3  $^{\pm0.10}$ & 0.06 $^{\pm0.00}$ &  2330 $^{\pm100}$ & 7.6 & --1091 & 1752 & 845~ & 4~~~\\
Dol-Dzim 10 &20	05 48 &40 32 00 &2.5  &0.07 &1.0 $^{\pm0.05}$  & 0.55 $^{\pm0.09}$ &  1670 $^{\pm75}$ & 8.3 & --384~ & 1620 & 135~ & 4~~~\\
Dol-Dzim 11 &20	51 00 &35 57 00 &3.0  &0.40 &2.0 $^{\pm0.11}$ & 0.42 $^{\pm0.05}$ &  2545 $^{\pm115}$ & 8.4 & --522~ & 2480 & --232~ & 4~~~\\
Ruprecht ~13 &07 08 03 &-25 52 00 &4.5 &0.07 &1.00 $^{\pm0.05}$ & 0.26 $^{\pm0.05}$ &  1300 $^{\pm60}$ & 9.3 & 685 & --1090 & --185 & 9~~~\\
Ruprecht ~15 &07 19 33 &-19 38 00 &5.5 &0.32 &0.50 $^{\pm0.03}$ & 0.65 $^{\pm0.05}$ &  1845 $^{\pm85}$ & 9.7 & 1095 & --1482 & --93 & 8~~~\\
Ruprecht ~16 &07 23 10 &-19 28 00 &3.5 &0.31 &0.16 $^{\pm0.02}$ & 0.71 $^{\pm0.07}$ &  2160 $^{\pm100}$ & 9.9 & 1276 & --1742 & --78 & 9~~~\\
Ruprecht ~24 &07 31 54 &-12 45 00 &5.0 &0.54 &0.06 $^{\pm0.00}$ & 0.35 $^{\pm0.05}$ &  1983 $^{\pm90}$ & 9.9 & 1300 & --1492 & 103 & 9~~~\\
Ruprecht 135 &17 58 12 &-11 39 00 &3.0 &0.50 &0.50 $^{\pm0.02}$ & 1.10 $^{\pm0.05}$ &  1850 $^{\pm85}$ & 6.74 & --1764 & 520 & 201 & 1~~~\\
Ruprecht 137 &18 00 16 &-25 13 39 &2.8 &0.25 &0.80 $^{\pm0.06}$ & 0.67 $^{\pm0.05}$ &  1450 $^{\pm65}$ & 7.06 & --1445 & 123 & --23 & 1~~~\\
Ruprecht 138 &17 59 56 &-24 40 57 &3.0 &0.21 &2.00 $^{\pm0.11}$ & 0.18 $^{\pm0.05}$ &  930 $^{\pm40}$  & 7.57 & --926 & 86 & --9 & 1~~~\\
Ruprecht 142 &18 32 11 &-12 13 47 &3.3 &0.03 &0.40 $^{\pm0.04}$ & 0.91 $^{\pm0.05}$ &  1735 $^{\pm80}$ & 6.89 & --1631 & 590 & --41 & 1~~~\\
Ruprecht 168 &17 52 46 &-28 26 00 &2.6 &0.80 &2.00 $^{\pm0.12}$ & 1.06 $^{\pm0.11}$ &  820 $^{\pm35}$ & 7.68 & --820 & 18 & --16 & 1~~~\\
Ruprecht 169 &17 59 22 &-24 46 01 &2.6 &0.14 &1.00 $^{\pm0.05}$ & 0.66 $^{\pm0.05}$ &  1390 $^{\pm60}$ & 7.12 & --1384 & 125 & --13 & 1~~~\\
Ruprecht 171 &18 32 11 &-16 02 59 &5.7 &0.26 &3.20 $^{\pm0.15}$ & 0.12 $^{\pm0.03}$ &  1140 $^{\pm50}$ & 7.41 & --1092 & 323 & --62 & 1~~~\\
Dolidze ~9  &20 25 42 &41 56 00 &3.0 &0.72 &0.02 $^{\pm0.00}$ & 0.80 $^{\pm0.05}$ &  ~866 $^{\pm40}$ & 8.4 & --152  & 852   & 35 & 5~~~\\
Dolidze 10  &20 26 18 &40 07 00 &1.8 &0.32 &0.25 $^{\pm0.05}$ & 0.80 $^{\pm0.05}$ &  ~950 $^{\pm44}$ & 8.4 & --190  & 931   & 19 & 5~~~\\
Dolidze 11  &20 26 30 &41 27 00 &2.5 &0.27 &0.40 $^{\pm0.03}$ & 0.83 $^{\pm0.06}$ &  1127 $^{\pm52}$ & 8.4 & --204  & 1108  & 37 & 5~~~\\
Dolidze 19  &05 23 42 &08 11 00 &12.0&0.40 &0.16 $^{\pm0.05}$ & 0.55 $^{\pm0.02}$ &  1320 $^{\pm60}$ & 9.8 & 1229   & --332 & --349 & 5~~~\\
Dolidze 21  &05 27 24 &07 04 00 &6.0 &0.65 &0.20 $^{\pm0.08}$ & 0.55 $^{\pm0.03}$ &  1447 $^{\pm65}$ & 9.9 & 1339   & --399 & --377 & 5~~~\\
Dolidze 26  &07 30 06 &11 54 00 &11.0&0.80 &0.10 $^{\pm0.01}$ & 0.44 $^{\pm0.04}$ &  1907 $^{\pm88}$ & 10.2& 1657   & --826 & 458 & 5~~~\\
Dolidze 27  &16 36 30 &-08 57 00&14.0&0.78 &0.05 $^{\pm0.00}$ & 0.75 $^{\pm0.07}$ &  1015 $^{\pm45}$ & 7.5 & --914  & 122   & 424 & 5~~~\\
Dolidze 37  &20 03 00 &37 41 00 &4.0 &0.40 &0.30 $^{\pm0.08}$ & 0.48 $^{\pm0.05}$ &  1490 $^{\pm70}$ & 8.2 & --411  & 1429  & 93 & 5~~~\\
Dolidze 39  &20 16 24 &37 52 00 &4.0 &0.06 &0.25 $^{\pm0.06}$ & 0.44 $^{\pm0.05}$ &  ~872 $^{\pm40}$ & 8.3 & --218  & 844   & 22 & 5~~~\\
Dolidze 41  &20 18 49 &37 45 00 &5.0 &0.25 &0.40 $^{\pm0.05}$ & 0.53 $^{\pm0.03}$ &  1763 $^{\pm80}$ & 8.2 & --435  & 1708  & 31 & 5~~~\\
\end{tabular}}
\end{table*}
\begin{table*}
\centerline{
\begin{tabular}{cccrrrrrrrrrr}
\hline
Cluster  & $\alpha$
            & $\delta$ & R$_{lim.}$ & R$_{c}$& Age  & E$_{B-V}$  & $R_{\odot}$  & $R_\mathrm{gc}$
            & $X_{\odot}$  & $Y_{\odot}$  & $Z$ & Ref.\\
            & $~~^{h}~~^{m}~~^{s}$ & $~~^{\circ}~~{'}~~{''}$ & ${'}$ & ${'}$ & \it Gyr & \it mag  & \it pc  & \it kpc   & \it pc
            & \it pc  & \it pc & \\
\hline
Dias 2  &06 09 09 &04 35 24  &5.5 &0.41 &0.79 $^{\pm0.02}$ & 0.61 $^{\pm0.11}$ &  2835 $^{\pm130}$ & 11.2 & 2570   & --1142 & --358 & 3~~~\\
Dias 3  &07 10 28 &-08 26 14 &8.0 &0.88 &1.41 $^{\pm0.11}$ & 0.64 $^{\pm0.11}$ &  4650 $^{\pm215}$ & 12.3 & 3423   & --3147 & 28 & 3~~~\\
Dias 4  &13 43 40 &-63 01 30 &3.2 &0.04 &1.26 $^{\pm0.09}$ & 0.60 $^{\pm0.10}$ &  2150 $^{\pm100}$ & 7.3  & --1347 & --1675 & --28 & 3~~~\\
Dias 6  &18 30 30 &-12 18 59 &3.0 &0.28 &0.60 $^{\pm0.02}$ & 0.91 $^{\pm0.07}$ &  1580 $^{\pm70}$ & 7.03 & --1488 & 530 & --28 & 1~~~\\
Dias 7  &19 49 22 &21 09 48  &5.0 &0.22 &2.00 $^{\pm0.10}$ & 0.42 $^{\pm0.04}$ &  2540 $^{\pm115}$ & 7.5  & --1334 & 2158   & --109 & 3~~~\\
Dias 8  &19 52 07 &11 37 54  &5.0 &0.12 &2.24 $^{\pm0.10}$ & 0.30 $^{\pm0.05}$ &  2220 $^{\pm100}$ & 7.3  & --1404 & 1693   & --302 & 3~~~\\
Turner ~2 &18 17 11 &-18 49 27 &3.8  &0.87 &0.10 $^{\pm0.01}$ & 0.36 $^{\pm0.04}$ &  1190 $^{\pm55}$ & 7.34 & --1163 & 250 & --27 & 1~~~\\
Turner ~6 &10 59 01 &-59 29 58 &1.3  &0.11 &0.08 $^{\pm0.01}$ & 1.32 $^{\pm0.13}$ &  3250 $^{\pm145}$ & 8.0 & 278 & --3071 & 18 & 4~~~\\
Turner ~7 &14 32 33 &-56 53 12 &13.0 &0.32 &0.08 $^{\pm0.01}$ & 1.39 $^{\pm0.15}$ &  1800 $^{\pm80}$ & 7.3 & --251 & --1238 & 104 & 4~~~\\
Turner ~8 &19 45 16 &27 50 30 &5.0  &0.34 &5.00 $^{\pm0.30}$ & 1.16 $^{\pm0.12}$ &  2160 $^{\pm100}$ & 7.8 & --30 & 1933 & 64 & 4~~~\\
Turner 11 &20 43 24 &35 35 18 &8.2  &0.45 &0.40 $^{\pm0.03}$ & 1.13 $^{\pm0.10}$ &  1905 $^{\pm85}$ & 8.3 & --30 & 1850 & --142 & 4~~~\\
King 17 &05 08 24 &39 05 00 &2.8 &0.16 &0.79 $^{\pm0.03}$ & 0.73 $^{\pm0.12}$ &  2960 $^{\pm135}$ & 11.4 & 2887   & 650    & --38 & 3~~~\\
King 18 &22 52 06 &58 17 00 &2.4 &0.16 &0.35 $^{\pm0.03}$ & 0.52 $^{\pm0.07}$ &  1860 $^{\pm85}$ & 9.2  & 567    & 1770   & --33 & 3~~~\\
King 23 &07 21 47 &-00 59 06 &3.6 &0.37 &0.89 $^{\pm0.04}$ & 0.16 $^{\pm0.05}$ &  3113 $^{\pm140}$ & 11.2 & 2513   & --1795 & 390 & 3~~~\\
King 26 &19 29 01 &14 52 02 &2.2 &0.24 &0.44 $^{\pm0.02}$ & 1.27 $^{\pm0.15}$ &  2600 $^{\pm120}$ & 7.1  & --1656 & 2003   & --61 & 3~~~\\
BH ~~47 &08 42 33 &-48 05 13 &7.7 &0.27 &0.80 $^{\pm0.05}$ & 0.73 $^{\pm0.05}$ &  2465 $^{\pm110}$ & 6.03 & 145 & --2456 & --154  & 1~~~\\
BH ~~60 &09 15 53 &-50 00 38 &2.1 &0.09 &0.40 $^{\pm0.01}$ & 0.67 $^{\pm0.05}$ &  1325 $^{\pm60}$ & 7.17 & --38 & --1324 & --16  & 1~~~\\
BH 218  &17 16 12 &-39 24 04 &2.8 &0.06 &0.40 $^{\pm0.01}$ & 0.88 $^{\pm0.06}$ &  1215 $^{\pm55}$ & 7.28 & --1188 & --254 & --14  & 1~~~\\
ESO 524-01    &18 56 37	&-26 57	39 &3.0 &0.30 &3.20 $^{\pm0.14}$ & 0.30 $^{\pm0.03}$ &  2800 $^{\pm130}$ & 5.75 & --2694 & 430 & --629 & 1~~~\\
ESO 522-05    &18 12 53	&-24 21 50 &2.2 &0.23 &3.20 $^{\pm0.14}$ & 1.82 $^{\pm0.15}$ &  660 $^{\pm30}$ & 7.84 & --654 & 80 & --35 & 1~~~\\
ESO 525-08    &19 27 16	&-23 34	35 &3.0 &0.26 &1.00 $^{\pm0.05}$ & 0.36 $^{\pm0.04}$ &  1640 $^{\pm75}$ & 6.93 & --1505 & 407 & --507 & 1~~~\\
IC 1434       &22 10 34	&52	49 40 &3.5 &0.27 &0.32 $^{\pm0.02}$ & 0.66 $^{\pm0.05}$ &  3035 $^{\pm140}$ & 9.5  & 523    & 2986   & --143 & 3~~~\\
IC 2156       &06 04 51	&24	09 30 &2.0 &0.07 &0.25 $^{\pm0.02}$ & 0.67 $^{\pm0.05}$ &  2100 $^{\pm95}$ & 10.6 & 2087   & --230  & 47  & 3~~~\\
IC 4291       &13 36 56	&-62 05	45&2.8 &0.09 &0.80 $^{\pm0.04}$ & 0.61 $^{\pm0.05}$ &  1790 $^{\pm82}$ & 7.53 & --1107 & --1406 & 10 & 1~~~\\
Alessi 15     &06 43 04 &01 40 19 &6.0 &0.12 &0.45 $^{\pm0.03}$ & 0.91 $^{\pm0.07}$ &  2509 $^{\pm116}$ & 10.74 & 2161 & --1273 & --48 & 7~~~\\
Alessi 53     &06 29 24 &09 10 39 &6.4 &0.13 &0.50 $^{\pm0.04}$ & 0.61 $^{\pm0.05}$ &  2360 $^{\pm109}$ & 10.72 & 2184 & --894 & --27 & 7~~~\\
Juchert ~1    &19 22 32 &12 40 00 &1.6 &0.04 &0.40 $^{\pm0.03}$ & 1.36 $^{\pm0.10}$ &  2286 $^{\pm105}$ & 7.16 & --1538 & 1691 & --40 & 7~~~\\
Juchert 12    &07 20 57 &-22 52 00 &5.0 &0.36 &0.30 $^{\pm0.02}$ & 0.91 $^{\pm0.07}$ &  3016 $^{\pm139}$ & 10.47 & 1658 & --2510 & --217 & 7~~~\\
Riddle ~4     &02 07 23 &60 15 25 &2.2 &0.13 &0.05 $^{\pm0.05}$ & 0.91 $^{\pm0.07}$ &  1993 $^{\pm92}$ & 9.95 & 1339 & 1475 & --43 & 7~~~\\
Riddle 15     &19 11 09 &14 50 04 &2.5 &0.08 &0.50 $^{\pm0.03}$ & 1.33 $^{\pm0.10}$ &  1925 $^{\pm89}$ & 7.36 & --1278 & 1437 & 82 & 7~~~\\
Skiff 1       &06 14 47 &12 52 15 &1.5 &0.10 &0.25 $^{\pm0.01}$ & 0.57 $^{\pm0.04}$ &  3150 $^{\pm145}$ & 5.35 & 3005 & --937 & --115 & 1~~~\\
Skiff 2       &04 58 14 &43 00 48 &2.5 &0.30 &0.90 $^{\pm0.06}$ & 0.18 $^{\pm0.05}$ &  2125 $^{\pm100}$ & 6.37 & 2032 & 621 & 5 & 1~~~\\
Teutsch ~11   &06 25 24 &13 51 59 &3.0 &0.26 &0.50 $^{\pm0.04}$ & 0.70 $^{\pm0.05}$ &  3443 $^{\pm159}$ & 11.83 & 3280 & --1044 & 39 & 7~~~\\
Teutsch 144   &21 21 44 &50 36 36 &5.0 &0.30 &0.80 $^{\pm0.08}$ & 0.73 $^{\pm0.05}$ &  1704 $^{\pm79}$ & 8.75 & 81 & 1702 & 14 & 7~~~\\
Collinder 351 &17 49 00 &-28 44 09 &4.2 &0.19 &0.16 $^{\pm0.02}$ & 0.70 $^{\pm0.05}$ &  1310 $^{\pm60}$ & 7.19 & --1310 & 14 & --16 & 1~~~\\
Patchick 89   &19 59 33 &49 18 45 &3.5 &0.07 &1.60 $^{\pm0.13}$ & 0.21 $^{\pm0.02}$ &  2646 $^{\pm122}$ & 8.62 & --288 & 2588 & 465 & 7~~~\\
Toepler 1     &20 01 18 &33 36 54 &4.0 &0.10 &0.40 $^{\pm0.03}$ & 0.79 $^{\pm0.05}$ &  2890 $^{\pm133}$ & 8.00 & --974 & 2719 & 87 & 7~~~\\
\hline\\
\end{tabular}}
\end{table*}
\begin{table*}
\renewcommand{\arraystretch}{0.80}
\caption[]{References of studied clusters.}
\begin{minipage}{\textwidth}
\begin{tabular}{ll} \hline

Number & Reference \\
\hline
1 & Tadross, A. L., 2008, New Astronomy 13, 370. \\
2 & Tadross, A. L., 2008, Monthly Notices of the Royal Astronomical Society 389, 285. \\
3 & Tadross, A. L., 2009, New Astronomy 14, 200. \\
4 & Tadross, A. L., 2009, Astrophys \& Space Sci. 323, 383. \\
5 & Tadross, A. L. \& Nasser, M. A., 2010, NRIAG Journal Ser."A67" (arXiv:1011.2934). \\
6 & Tadross, A. L., 2011, Journal of the Korean Astronomical Society 44 (1), 1. \\
7 & Tadross, A. L., et al, 2012, Research in Astronomy and Astrophysics (RAA) 12, 75. \\
8 & Tadross, A. L., 2012, Research in Astronomy and Astrophysics (RAA) 12, 158. \\
9 & Tadross, A. L., 2012, New Astronomy 17, 198. \\
\hline \label{OCdata}
\end{tabular}
\end{minipage}
\end{table*}

\end{document}